\documentclass[trackchanges, twocolumn]{aastex701}

\shorttitle{Panchromatic JWST Transmission Spectrum of LTT 9779 ${\rm b}$}
\shortauthors{Stamer et al.}

\usepackage[version=4]{mhchem}
\usepackage{amsmath}
\usepackage{caption}
\usepackage{subcaption}
\usepackage{graphicx}
\usepackage{booktabs}
\usepackage{gensymb}
\usepackage{array}
\usepackage{caption}
\usepackage{rotating} 
\usepackage{placeins}

\DeclareMathSymbol{\mh}{\mathord}{operators}{`\-}

\usepackage{tabularx}
\usepackage{comment}
\usepackage{wrapfig}
\usepackage{float}
\begin{document}

\title{Cool, Cloudy Nights on an Ultrahot Neptune: \\
A Panchromatic NIRISS/NIRSpec Transmission Spectrum of LTT 9779 b}

\author[0000-0003-1131-8922]{Sarah Stamer}
\affiliation{Department of Physics and Astronomy, University of New Mexico, Albuquerque, NM 87106, USA}
\email{sstamer@unm.edu}  
\correspondingauthor{Sarah Stamer, sstamer@unm.edu}

\author[0000-0003-2313-467X]{Diana Dragomir} 
\affiliation{Department of Physics and Astronomy, University of New Mexico, Albuquerque, NM 87106, USA}
\email{dragomir@unm.edu}

\author[0000-0002-7500-7173]{Annabella Meech} 
\affiliation{Space Telescope Science Institute, 3700 San Martin Dr., Baltimore, MD 21218, USA}
\email{ameech@stsci.edu}

\author[0000-0003-3667-8633]{Joshua D. Lothringer} 
\affiliation{Space Telescope Science Institute, 3700 San Martin Dr., Baltimore, MD 21218, USA}
\email{jlothringer@stsci.edu}

\author[0000-0002-0786-7307]{Reza Ashtari} 
\affiliation{Johns Hopkins University Applied Physics Laboratory, Laurel, MD 20723, USA}
\email{Reza.Ashtari@jhuapl.edu}

\author[0000-0002-2072-6541]{Jonathan Brande} 
\affiliation{Department of Physics and Astronomy, University of Kansas, Lawrence, KS 66045, USA}
\affiliation{University of Maryland, 7901 Regents Drive, College Park, MD 207426} 
\email{jbrande@umd.edu}

\author[0000-0001-9521-6258]{Vivien Parmentier} 
\affiliation{Universite Cote d'Azur, Av. Valrose, 06000 Nice, France}
\email{vivien.parmentier@oca.eu} 

\author{Ian J. M. Crossfield} 
\affiliation{Department of Physics and Astronomy, University of Kansas, Lawrence, KS 66045, USA}
\email{ianc@ku.edu} 

\author[0000-0002-6939-9211]{Tansu Daylan}
\affiliation{Department of Physics and McDonnell Center for the Space Sciences, Washington University, St. Louis, MO 63130, USA}
\email{tansu@wustl.edu}

\author[0000-0001-8018-0264]{Suman Saha} 
\affiliation{Instituto de Estudios Astrofísicos, Facultad de Ingeniería Ciencias, Universidad Diego Portales, Santiago, Chile}
\affiliation{Centro de Excelencia en Astrofísica y Tecnologías Afines (CATA), Las Condes, Santiago, Chile}
\email{suman.saha@mail.udp.cl}

\author[0000-0001-7714-7551]{K. Angelique Kahle} 
\affiliation{Max Planck Institute for Astronomy, 69117 Heidelberg, Germany}
\affiliation{Department of Physics and Astronomy, Heidelberg University, 69120 Heidelberg, Germany}
\email{kahle@mpia.de}

\author[0000-0002-7352-7941]{Kevin B. Stevenson} 
\affiliation{Johns Hopkins University Applied Physics Laboratory, Laurel, MD 20723, USA}
\email{Kevin.Stevenson@jhuapl.edu}

\author[0000-0002-1221-5346]{David R. Coria} 
\affiliation{Department of Physics and Astronomy, University of New Mexico, Albuquerque, NM 87106, USA}
\email{drcoria@unm.edu}

\author[0000-0001-9987-467X]{Jared E.J. Splinter}
\affiliation{Trottier Space Institute at McGill, 3550 rue University, Montr\'eal, QC H3A 2A7, Canada}
\affiliation{Department of Earth and Planetary Sciences, McGill University, 3450 rue University, Montr\'eal, QC H3A OE8, Canada}
\email{jared.splinter@mail.mcgill.ca}

\author[0000-0001-6129-5699]{Nicolas B. Cowan}
\affiliation{Department of Physics, 3600 rue University, Montr\'eal, QC H3A 2T8, Canada}
\affiliation{Department of Earth and Planetary Sciences, McGill University, 3450 rue University, Montr\'eal, QC H3A OE8, Canada}
\email{nicolas.cowan@mcgill.ca}

\author[0000-0002-2341-3233]{Emma Esparza-Borges}
\affiliation{European Space Agency (ESA), European Space Astronomy Centre (ESAC), Camino Bajo del Castillo s/n, 28692 Villanueva de
la Ca{\~n}ada, Madrid, Spain}
\email{emma.esparza-borges@esa.int}

\begin{abstract}

Studies of planets that reside in the hot Neptune desert allow us to gain insight into how their atmospheres survive the intense flux from their host stars. One such member of that population, LTT 9779 b, is unique in that it has a confirmed atmosphere across various multi-wavelength studies, although results regarding the planet's atmospheric content differ. We present a panchromatic ($\lambda = 0.63 - 5.165$ $\mu$m) JWST transmission spectrum of LTT 9779 b's atmosphere, including a re-reduction of previously obtained NIRISS/SOSS data and a reduction of recent NIRSpec/G395H data. Through atmospheric retrievals, we find muted spectral features consistent with the presence of clouds and a low terminator temperature ($\sim560 \ \mathrm{K}$ for free chemistry, and $\sim910-950 \ \mathrm{K}$ for equilibrium chemistry), with a range of possible atmospheric metallicity values dependent on atmospheric temperature and cloud coverage. Forward modeling suggests that, if clouds are present, small sub-micron particle sizes, moderate-to-high sedimentation efficiency ($f_{\rm sed} = 1 - 3$, corresponding to a relatively compact vertical cloud extent), low particle column densities, and high eddy diffusion coefficients ($k_{zz} \geq 10^{10}$ cm$^{2}$s$^{-1}$) are preferred. Probing silicate cloud absorption features in the mid-infrared is needed to confirm the cloud species, allowing us to break the cloud-metallicity degeneracy that is present in this panchromatic transmission spectrum. 

\end{abstract}


\section{Introduction}\label{sec:intro}

With the growing number of exoplanets comes the ability to explore demographic features and to probe new and unique populations of planets, for example, planets located close to their host stars. One particularly interesting demographic feature is the hot Neptune desert \citep{2011ApJ...727L..44S, 2016A&A...589A..75M}, where there is a lack of Neptune-mass planets (or planets with radii between $\sim$2 $R_{\oplus}$ and 10 $R_{\oplus}$) with orbital periods shorter than $\sim$3 days. The Hot Neptune desert, or evaporation desert \citep{2014ApJ...787...47S, 2016NatCo...711201L}, is believed to result from the process of photoevaporation, in which intense high-energy radiation from a planet’s host star, particularly in the X-ray and extreme ultraviolet (XUV) range, heats the upper atmosphere and drives atmospheric escape. Neptune-sized planets that form or migrate into very close-in orbits are subjected to this extreme irradiation, which can remove their H-/He-rich envelopes over relatively short timescales \citep{2012MNRAS.425.2931O, 2017MNRAS.472..245L}. These planets typically lack the gravitational binding energy required to retain their atmospheres under such conditions, making them especially vulnerable to rapid mass loss, particularly during the first 100 million years of their evolution when stellar activity is at its peak \citep{2013ApJ...775..105O, 2018MNRAS.479.5012O}. In contrast, more massive gas giants can retain their atmospheres, while smaller rocky planets may have already lost their atmospheres or never accreted significant gaseous envelopes in the first place. 

\begin{figure}
    \centering
    \includegraphics[width=\columnwidth]{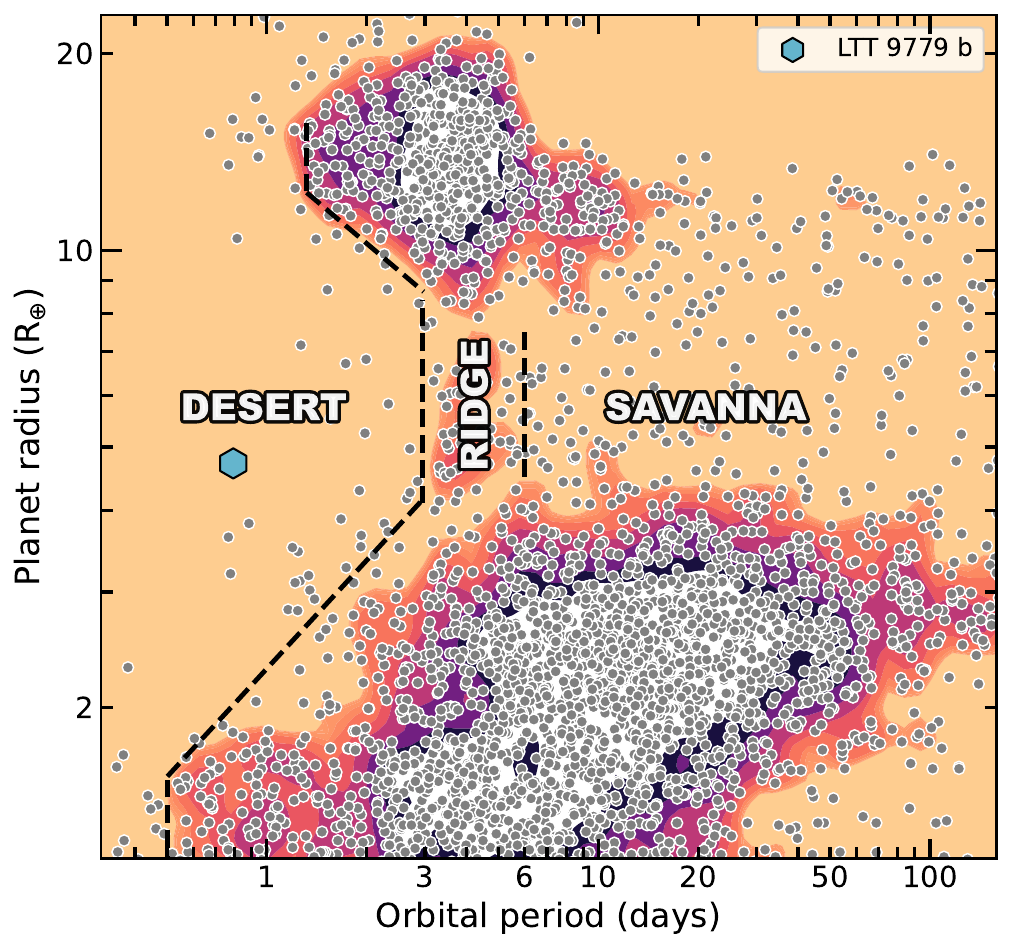}
    \caption{Location of LTT 9779 b in radius-period space. The background population and boundaries of the hot Neptune desert are adapted from \citealt{2024A&A...689A.250C}, with LTT 9779 b highlighted for comparison.}
    \label{fig:nep_desert}
\end{figure}

LTT 9779 b, which was discovered by the Transiting Exoplanet Survey Satellite (TESS; \citealt{2015JATIS...1a4003R}), is one of a small population of known exoplanets that reside in the hot Neptune desert (Figure \ref{fig:nep_desert}), alongside planets such as NGTS-4 b, TOI-132 b, TOI-2196 b, and TOI-1853 b \citep{2019MNRAS.486.5094W, 2020MNRAS.493..973D, 2022A&A...666A.184P, 2023Natur.622..255N}. With a mass of 29.32 $M_{\oplus}$ (1.71 Neptunian masses) and a radius of 4.72 $R_{\oplus}$, the planet's density is similar to that of Neptune \citep{2020NatAs...4.1148J}. However, its orbital period of 0.792 days (or $\sim$19 hours) places it well within the hot Neptune desert and below the $\sim$1 day period cutoff for this planet to be considered an ultrahot Neptune \citep{2013ApJ...774...54S, 2018NewAR..83...37W}, as reflected in its equilibrium temperature of 1978 $\pm$ 19 K \citep{2020NatAs...4.1148J}. LTT 9779 b is one of only a few ultrahot Neptunes discovered thus far, alongside other rare ultra-short-period Neptune-sized planets such as TOI-849 b, TOI-332 b, and TOI-3261 b \citep{2020Natur.583...39A, 2023MNRAS.526..548O, 2024AJ....168..132N}. Unlike several of these highly irradiated planets, LTT 9779 b is a mature, low-density planet that has retained a substantial atmosphere, making it a high-priority target for transmission spectroscopy \citep{2020NatAs...4.1148J} and other atmospheric studies.

Through prior secondary eclipse, phase curve, transmission, and high-resolution spectroscopic studies, we have been able to confirm and increasingly characterize the atmosphere of this unique ultrahot Neptune.

Secondary eclipse observations first established that LTT 9779 b has retained a substantial atmosphere and subsequently revealed its unusually reflective nature. Secondary eclipse measurements at 3.6 and 4.5 $\mu$m from Spitzer/IRAC, when combined with a secondary eclipse from TESS, found evidence for spectral features including molecular absorption, with atmospheric modeling preferring \ce{CO} over \ce{CO2} and \ce{H2O}, and provided the first evidence that the planet has retained its atmosphere \citep{2020ApJ...903L...6D}. Analysis of CHEOPS secondary eclipse measurements later yielded a high geometric albedo of 0.8 for LTT 9779 b, accounting for both reflected light and thermal emission. Combining these eclipse measurements with those from Spitzer and TESS placed a lower limit on the atmospheric metallicity and inferred the presence of silicate clouds \citep{2023A&A...675A..81H}. HST/WFC3 UVIS G280 (0.2--0.8 $\mu$m) eclipse injection-recovery tests further showed that silicate clouds likely explain the planet's high dayside albedo \citep{2025MNRAS.538.1853R}. More recently, an updated set of transit and ephemeris parameters was obtained through analysis of three TESS sectors and applied to 20 CHEOPS secondary eclipse measurements, producing a more precise geometric albedo of $\sim$0.73 and again finding evidence for silicate clouds \citep{2025A&A...700A..45S}.

Phase curve observations have provided complementary constraints on the planet's atmospheric composition and longitudinal cloud structure. Early phase curve measurements suggested that LTT 9779 b possesses a high-metallicity atmosphere \citep{2020ApJ...903L...7C}. More recently, phase-resolved observations revealed an asymmetric dayside, with a highly reflective western hemisphere ($A=0.79$) associated with clouds, a lower-albedo eastern hemisphere ($A=0.41$), and an overall dayside albedo of 0.50 when considering reflected light alone \citep{2025NatAs...9..512C}. These observations indicate substantial longitudinal variation in the cloud properties of the atmosphere.

Transmission spectroscopy has provided additional constraints on the atmospheric composition, cloud structure, and possible escape. Transit observations with HST/WFC3 G102 and G141 (0.8--1.6 $\mu$m) were best fit by an atmospheric model containing \ce{H2O}, \ce{CO2}, and \ce{FeH}, while observations of the 1.083 $\mu$m \ce{He} line showed no evidence for an escaping atmosphere \citep{2023AJ....166..158E}. A subsequent re-analysis of these data retrieved higher \ce{OH} abundances than \ce{H2O} abundances but did not find statistically significant evidence for the presence of any individual molecular species \citep{2025MNRAS.540..650Z}. Transmission spectroscopy with JWST/NIRISS (0.6--2.85 $\mu$m) similarly revealed muted spectral features, with retrievals preferring silicate clouds at millibar pressures. However, the NIRISS spectrum alone was insufficient to distinguish between a \ce{CH4}-dominated and an \ce{H2O}-dominated atmosphere \citep{radica2024muted}.

Ground-based spectroscopy has provided complementary constraints on both atmospheric composition and cloud properties. Transmission spectroscopy with VLT/ESPRESSO (0.4--0.78 $\mu$m) did not reveal a statistically significant atmospheric signal but placed a lower limit on the atmospheric metallicity that is consistent with the lack of detectable absorption features \citep{2025A&A...695A..26R}. Through ESPRESSO reflected-light high-resolution cross-correlation spectroscopy, TiO depletion has been suggested in the planet's western hemisphere, and upper limits have been placed on the cloud-top pressures on both the western and eastern sides of the planet, consistent with the cloud altitudes inferred from NIRISS/SOSS observations \citep{2026A&A...705A..27V}. A re-analysis of the ESPRESSO high-resolution cross-correlation spectrum subsequently reported a detection of reflected light from the planet and a high albedo of 0.88, providing further evidence for a highly reflective atmosphere \citep{2026A&A...711L...5B}.

Independent observations have also strengthened the case that LTT 9779 b has retained its atmosphere despite its extreme irradiation. The 1.083 $\mu$m \ce{He} line was observed with the WINERED spectrograph on the Magellan Clay Telescope, resulting in another non-detection of atmospheric escape, which the authors suggested may be related to the planet's high atmospheric metallicity \citep{2024ApJ...962L..19V}. In addition, XMM-Newton observations showed that the host star (spectral type 	G7 V) has an X-ray luminosity approximately a factor of 15 lower than expected, reducing the high-energy irradiation available to drive atmospheric mass loss \citep{2024MNRAS.527..911F}. Taken together, these observations consistently point toward a metal-rich, cloudy, and highly reflective atmosphere that has survived within the hot Neptune desert.

This work aims to build upon previous observations of this planet by studying near-IR transmission at longer wavelengths than those probed by NIRISS up to wavelengths explored in the early Spitzer studies. In addition, this work aims to further constrain the planet's atmospheric metallicity. Previous studies have primarily pointed to supersolar metallicity for LTT 9779 b's atmosphere \citep{2020ApJ...903L...7C, 2023A&A...675A..81H, 2024ApJ...962L..19V, 2025A&A...695A..26R, 2025A&A...700A..45S, 2026AJ....171..215A}. However, the possibility of the atmosphere having a subsolar metallicity is also present in the literature, primarily due to degeneracies between clouds and metallicity \citep{2023AJ....166..158E, radica2024muted, 2025NatAs...9..512C}.

We present a transmission spectrum combining data obtained using the James Webb Space Telescope (JWST) Near-Infrared Spectrograph (NIRSpec; \citealt{jakobsen2022}) G395H in the Bright Object Time Series (BOTS; \citealt{birkmann2022}) mode, and the JWST Near Infrared Imager and Slitless Spectrograph (NIRISS; \citealt{doyon2023near}) instrument in Single Object Slitless Spectroscopy (SOSS; \citealt{albert2023near}) mode. This panchromatic, high-precision transmission spectrum is particularly useful for addressing the cloud-metallicity degeneracy, as the broad wavelength coverage helps constrain the influence of clouds while extending to wavelengths that probe strong molecular features such as \ce{CO2}, an important tracer of atmospheric metallicity.

Additional studies of LTT 9779 b from this observing program are presented in \citealt{2026AJ....171..215A} (NIRSpec phase-curve observations), \citealt{2026ApJ..1005L..79S} (panchromatic NIRISS and NIRSpec secondary-eclipse observations), and \citealt{Brande2026} (NIRSpec secondary-eclipse observations).

\section{Observations and Data Reduction}\label{sec:obs}
We extracted the transit data from NIRISS and NIRSpec full-orbit phase curves, and used \texttt{Eureka!} \citep{eureka} for data reduction to obtain transmission spectra. 

\subsection{Observations}
We observed a full-orbit phase curve, including two secondary eclipses and one transit, with JWST NIRSpec/G395H. These observations were conducted as part of the GO 3231 program (PI: Crossfield). Observations occurred from October 17, 2024, 07:35:52 UT to October 18, 2024, 05:10:30 UT (total observation time: 24.97 hours). The transit occurred in the middle segment (out of 7) of the data for the entire full-orbit observation. We utilized data from both NIRSpec detectors, NRS1 and NRS2. 

We also reduced additional full-orbit phase curve data obtained with JWST NIRISS/SOSS, which was obtained as part of GTO 1201 (PI: Lafrenière) and presented previously in \citealt{radica2024muted} and \citealt{2025NatAs...9..512C} (Section \ref{subsec:NIRISS_vs_Radica} contains a comparison of the NIRISS transmission spectrum re-reduction in this work with the reduction from \citealt{radica2024muted}). We reduced the data from the entire observation, but proceeded with only the transit portion of the light curve, including some pre- and post-transit baseline, for the next steps of the analysis. We utilized data from the 1st and 2nd NIRISS orders. 

The data described here may be obtained from the MAST archive at \dataset[https://doi.org/10.17909/4wvw-hn83]{https://doi.org/10.17909/4wvw-hn83}. 

\subsection{NIRSpec Data Reduction}\label{subsec:NIRSpec_reduction}

We reduced the segment of transit data for NRS1 and NRS2, created spectroscopic and white-light curves, and obtained a final NIRSpec transmission spectrum using the \texttt{Eureka!} pipeline \citep{eureka}, v. 1.1. Stages 1 and 2 act as a wrapper for the \texttt{jwst} pipeline \citep{jwstpipeline}. Stage 3 performs background subtraction and extracts the optimal spectrum. Stage 4 generates the spectroscopic and white-light curves. Stage 5 fits the light curves, and Stage 6 generates the planetary transmission spectrum.

For Stage 1, all detector-level calibration steps were performed except for the dark-current correction, which was omitted because the NIRSpec detectors have a low dark current and this contribution is negligible for these bright-object time-series observations. The sigma threshold for jump detection\footnote{\url{https://www.stsci.edu/files/live/sites/www/files/home/jwst/documentation/technical-documents/_documents/JWST-STScI-008975.pdf}} was 4 for NRS1 and 7 for NRS2. No bias correction was applied to remain consistent with the \texttt{jwst} pipeline. Median group-level background subtraction was performed before ramp fitting. The curved trace was masked using a smoothing length of 11 pixels for the trace location to smooth over individual bad pixels while following the shape of the spectral trace, an aperture of 8 pixels around the trace to capture the source flux while leaving sufficient background on both sides, and, for NRS1, columns below an index of 650 were excluded when constructing the trace mask.

For Stage 2, we ran the default steps. We used the CRDS context pmap 1364 for NRS1 and 1303 for NRS2.

For Stage 3, we first masked NaNs and infs in the data arrays, which resulted in masking less than 0.7\% of all pixels. The source position was computed using a Gaussian fit. Outlier rejection in time was performed using the full frame with a double-iteration 4-sigma threshold (\texttt{Eureka!}'s default value), flagging less than 0.1\% of all pixels. We then performed column-by-column background subtraction. For each detector column, corresponding approximately to a wavelength channel, we estimated a constant background level, with outlying background pixels rejected using the median absolute deviation. The background region was defined using a half-width of 4 pixels for NRS1 and 6 pixels for NRS2 relative to the source position. For spectral extraction, we used an aperture half-width of 2 pixels for NRS1 and 4 pixels for NRS2, both relative to the source position. These values were chosen to maximize photometric precision. We used the median frame to construct the normalized spatial profile for optimal spectral extraction, with a smoothing window length of 13.

\begin{figure*}[t!]
\centering

\begin{minipage}[t]{\columnwidth}
    \centering

    \begin{subfigure}[t]{\linewidth}
        \centering
        \includegraphics[
            width=\linewidth
        ]{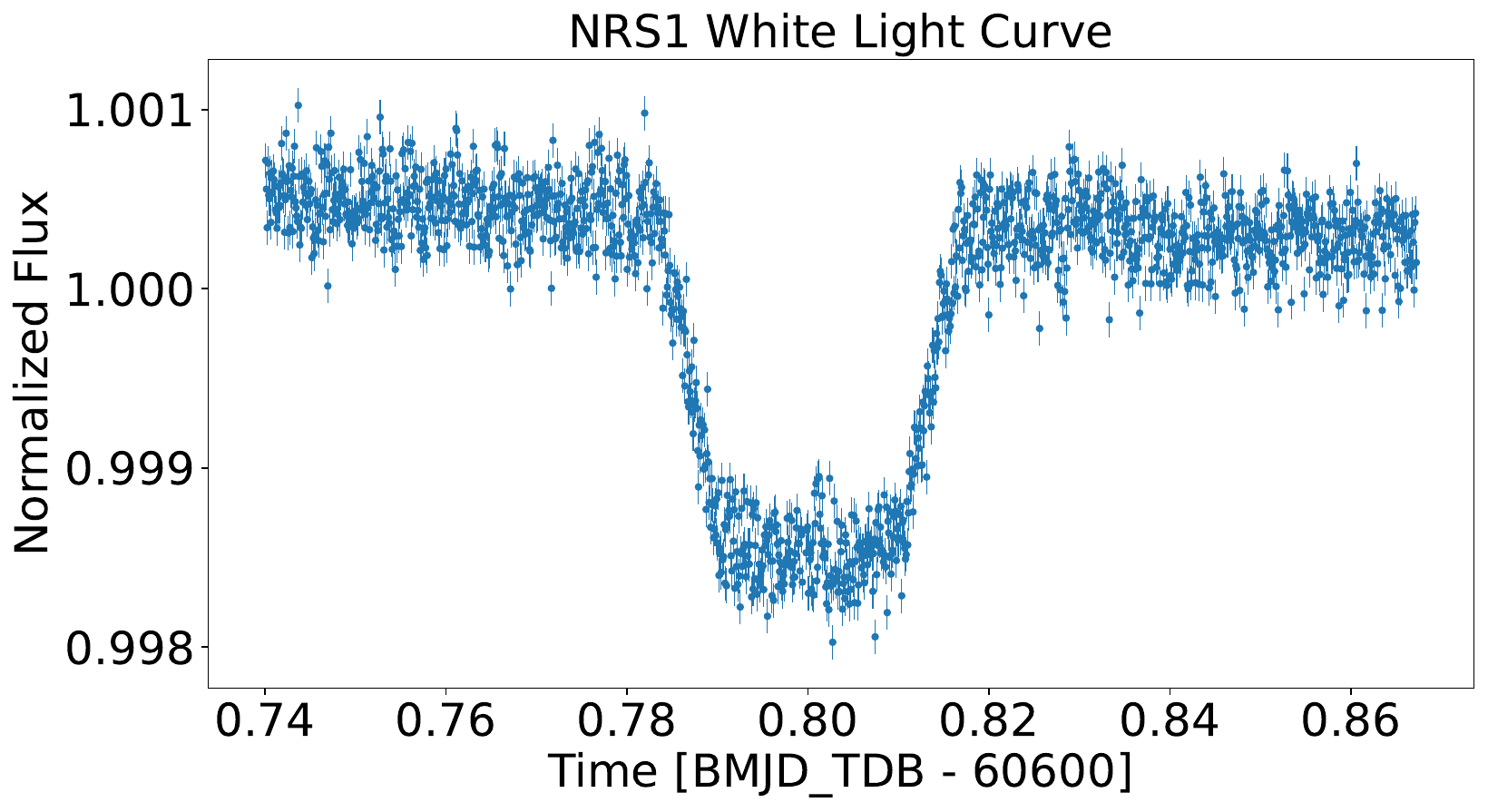}
    \end{subfigure}

    \vspace{0.2em}

    \begin{subfigure}[t]{\linewidth}
        \centering
        \includegraphics[
            width=\linewidth
        ]{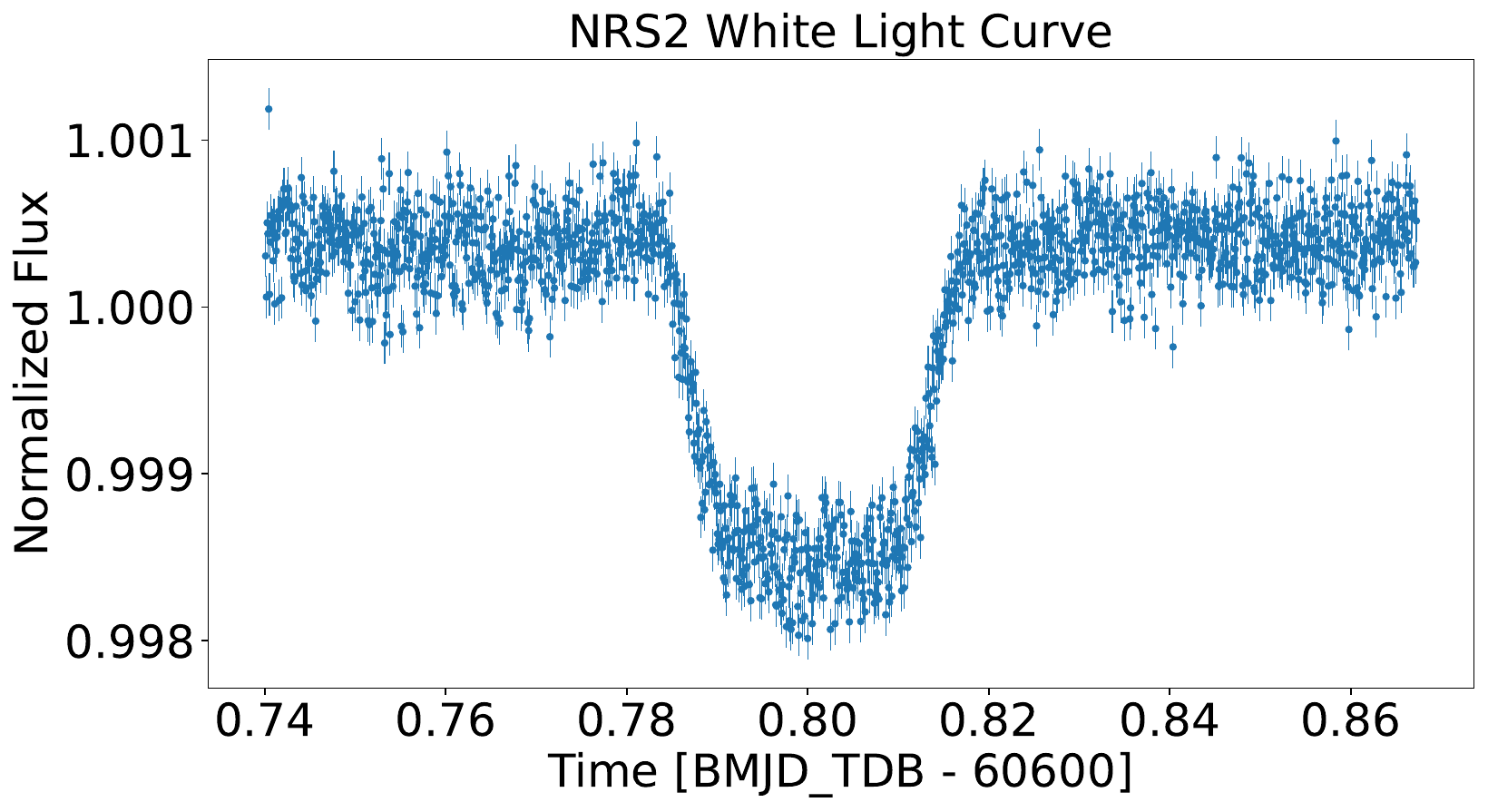}
    \end{subfigure}

    \caption{Broadband JWST/NIRSpec G395H white-light curves for the NRS1 and NRS2 detectors reduced with \texttt{Eureka!}. The points show the normalized stellar flux as a function of time across the transit for each detector.}
    \label{fig:NIRSpec_WLC}
\end{minipage}
\hfill
%
\begin{minipage}[t]{\columnwidth}
    \centering

    \begin{subfigure}[t]{\linewidth}
        \centering
        \includegraphics[
            width=\linewidth
        ]{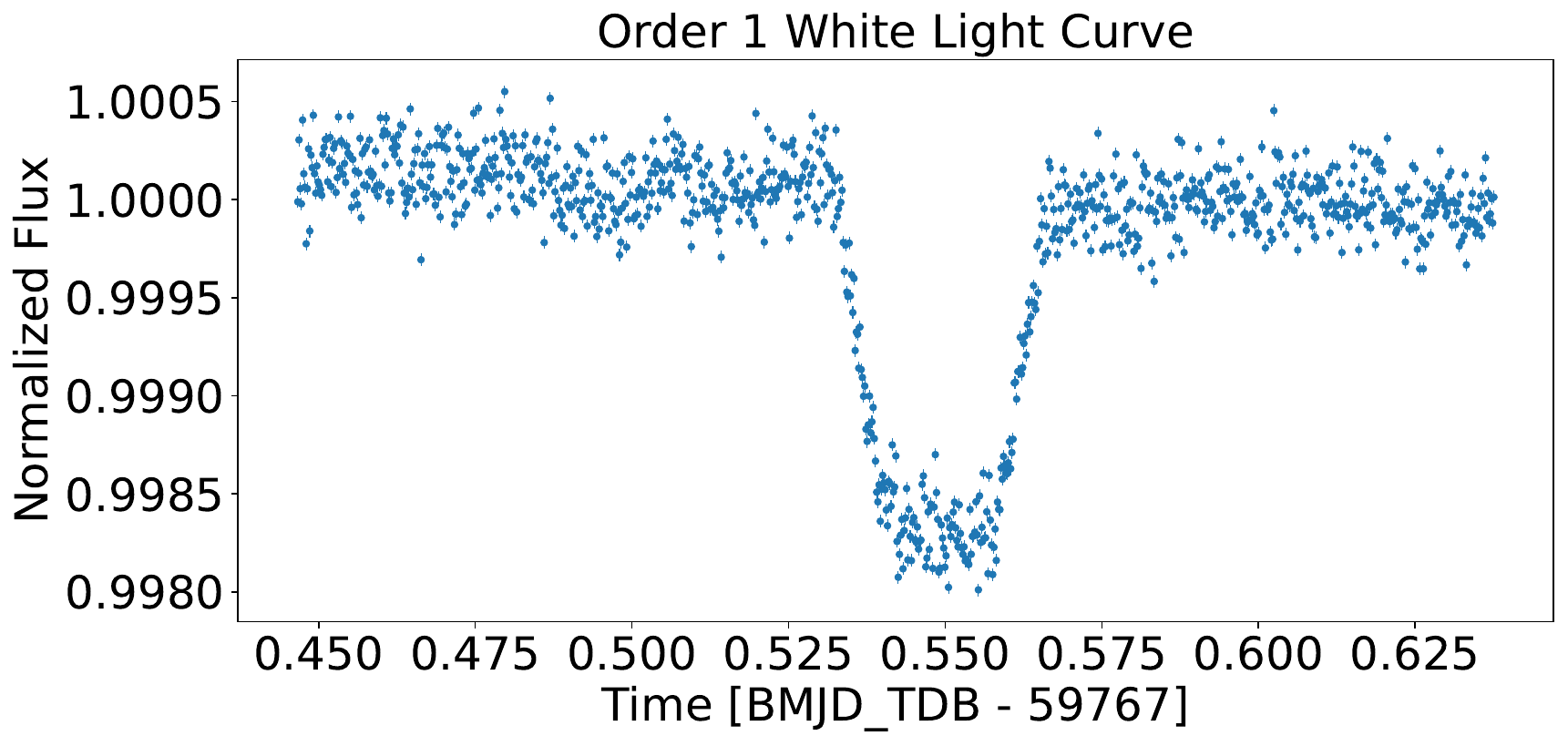}
    \end{subfigure}

    \vspace{0.2em}

    \begin{subfigure}[t]{\linewidth}
        \centering
        \includegraphics[
            width=\linewidth
        ]{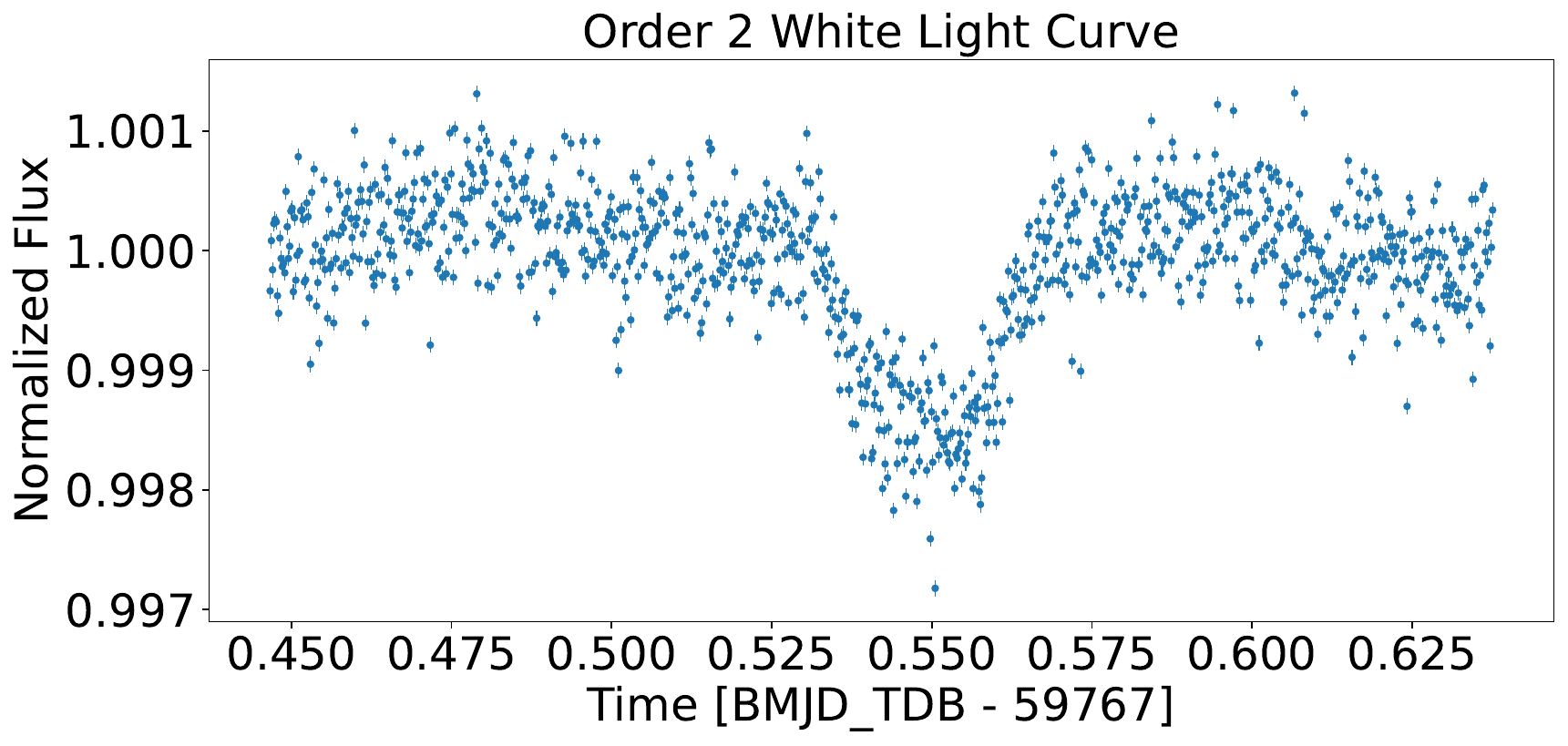}
    \end{subfigure}

    \caption{Broadband JWST/NIRISS SOSS white-light curves for Orders 1 and 2 reduced with \texttt{Eureka!}. The points show the normalized stellar flux as a function of time across the transit for each spectral order.}
    \label{fig:NIRISS_WLC}
\end{minipage}

\end{figure*}

For Stage 4, we computed spectroscopic and white-light curves for NRS1 and NRS2. The adopted white-light bandpasses were 2.817--3.720 $\mu$m for NRS1 and 3.823--5.165 $\mu$m for NRS2. The shortest-wavelength NRS1 channels were excluded because of increased noise near the detector edge. The resulting white-light curves are shown in Figure \ref{fig:NIRSpec_WLC}. We performed sigma-clipping on the binned 1D time series using a 7-sigma outlier threshold and a 30-point width for the box-car filter used to calculate the rolling median. Points flagged by this procedure were masked and therefore excluded from the Stage 5 light-curve fitting. We used 48 spectral channels for NRS1 and 70 spectral channels for NRS2. The resulting spectral resolutions are $R \approx 350$ for NRS1 and $R \approx 470$ for NRS2. We used \texttt{ExoTiC-LD} \citep{2024JOSS....9.6816G} to compute wavelength-dependent quadratic limb-darkening parameters $u_1$ and $u_2$ with the Stagger grid \citep{2015A&A...573A..90M}, using the stellar metallicity, effective temperature, and surface gravity published in \citealt{2020NatAs...4.1148J}. We also tested free limb-darkening coefficients; the procedure used to obtain these coefficients, a comparison with the \texttt{ExoTiC-LD} coefficients, and the corresponding NIRSpec spectra are discussed in Section \ref{subsec:LD_comparison}.

For Stage 5, we used the NRS1 and NRS2 white-light curves to refine the orbital parameters adopted in the subsequent spectroscopic light-curve fits. The priors and fitting parameters used for the white-light analysis are summarized in Table \ref{tab:lc_fit}. We fit $t_0$, the mid-transit time, as a shared value between the two NIRSpec light curves with a uniform prior, while $i$ and $a/R_*$ were treated as shared parameters with priors based on the updated TESS transit parameters from \citealt{2025A&A...700A..45S}. The fixed orbital period, $P=0.792064182$~d, and the adopted prior values for $i$ and $a/R_*$ were also taken from \citealt{2025A&A...700A..45S}.

We then performed the spectroscopic light-curve fits using the $t_0$ value from the joint white-light curve fits (60600.8 BMJD for NIRSpec), together with the mean values $i=76.0248^\circ$ and $a/R_*=3.8011$ obtained from the joint white-light analysis. The wavelength-dependent $u_1$ and $u_2$ limb-darkening coefficients calculated with \texttt{ExoTiC-LD} in Stage 4 were adopted for the spectroscopic fits. For each spectroscopic light curve, the free parameters were $R_\mathrm{p}/R_*$, $c_0$, $c_1$, and a multiplier to the expected noise level. The parameters $c_0$ and $c_1$ describe a first-order polynomial used to account for the out-of-transit flux level and a linear trend in the baseline.

We used \texttt{emcee} \citep{Foreman-Mackey2013} as the fitting method, with 2500 steps, 200 walkers, and 500 burn-in steps. The transit was modeled using \texttt{batman} \citep{Kreidberg2015} together with the first-order polynomial systematics model.

For each detector/order, we additionally calculated an inverse-variance weighted mean of the $R_\mathrm{p}/R_*$ values from the retained spectroscopic channels. These values are reported in Table \ref{tab:lc_fit} as representative channel-weighted radius ratios and should not be interpreted as broadband white-light-fit values.

We obtained the final \texttt{Eureka!} transmission spectrum in Stage 6.

\subsection{NIRISS Data Reduction}\label{subsec:NIRISS_reduction}

To reduce the transit data for SOSS Orders 1 and 2 and obtain the NIRISS transmission spectrum, we used v. 1.2.1 of the \texttt{Eureka!} pipeline, as it offers initial NIRISS reduction support. The data reduction process was similar to that used for NIRSpec.

Differences include the following. In Stage 2, we did not skip flat-fielding. Since SOSS is slitless, the pixel-to-pixel sensitivity variations are not self-calibrated, so this correction is necessary. We used the CRDS context pmap 1364 for both orders. In Stage 3, the background half-width was 22 pixels, and the half-width of the aperture region for spectral extraction was 17 pixels, both relative to the source position, with these values tuned to capture flux from the PSF without contamination from the overlapping orders or background sources. The smoothing window length was 7, chosen to suppress bad-pixel noise while tracking the real trace curvature.

For Stage 4, the adopted white-light bandpasses were 0.85--2.735 $\mu$m for SOSS Order 1 and 0.63--0.85 $\mu$m for SOSS Order 2. We performed sigma-clipping on the binned 1D time series using a 3-sigma outlier threshold, which is more aggressive than the NIRSpec threshold of 7 sigma while remaining appropriate for the SOSS noise properties, and a 31-point width for the box-car filter to span enough integrations to define a stable rolling-median baseline. Order 1 had 100 spectral channels and Order 2 had 12 spectral channels, again using approximately constant resolving power, resulting in $R \approx 190$ for Order 1 and $R \approx 80$ for Order 2.

In Stage 5, we applied a manual clip to isolate the transit while retaining pre- and post-transit baseline. The clipped Stage 4 white-light curves are shown in Figure \ref{fig:NIRISS_WLC}. The spectroscopic light curves were then fit using the same general approach as for NIRSpec, with $R_\mathrm{p}/R_*$ and the baseline/systematics parameters allowed to vary for each wavelength channel.

We note increased structure in the NIRISS spectrum near $\sim2.4\,\mu\mathrm{m}$. This structure is substantially less pronounced in the independent reduction of the same NIRISS observations presented by \citealt{radica2024muted} (Section \ref{subsec:NIRISS_vs_Radica}). We therefore do not interpret this feature as robust evidence for an atmospheric spectral feature and caution that it may reflect reduction-dependent systematics toward the long-wavelength end of the NIRISS/SOSS Order 1 spectrum.

The adopted light-curve fitting parameters are summarized in Table \ref{tab:lc_fit}. For consistency with the NIRSpec analysis, Table \ref{tab:lc_fit} also reports the inverse-variance weighted mean $R_\mathrm{p}/R_*$ obtained from the retained spectroscopic channels for each SOSS order; these values are not broadband white-light-fit parameters.

\begin{figure*}[htb!]
    \centering

    \includegraphics[
        width=\linewidth,
        trim={0cm 0cm 0cm 0cm},
        clip
    ]{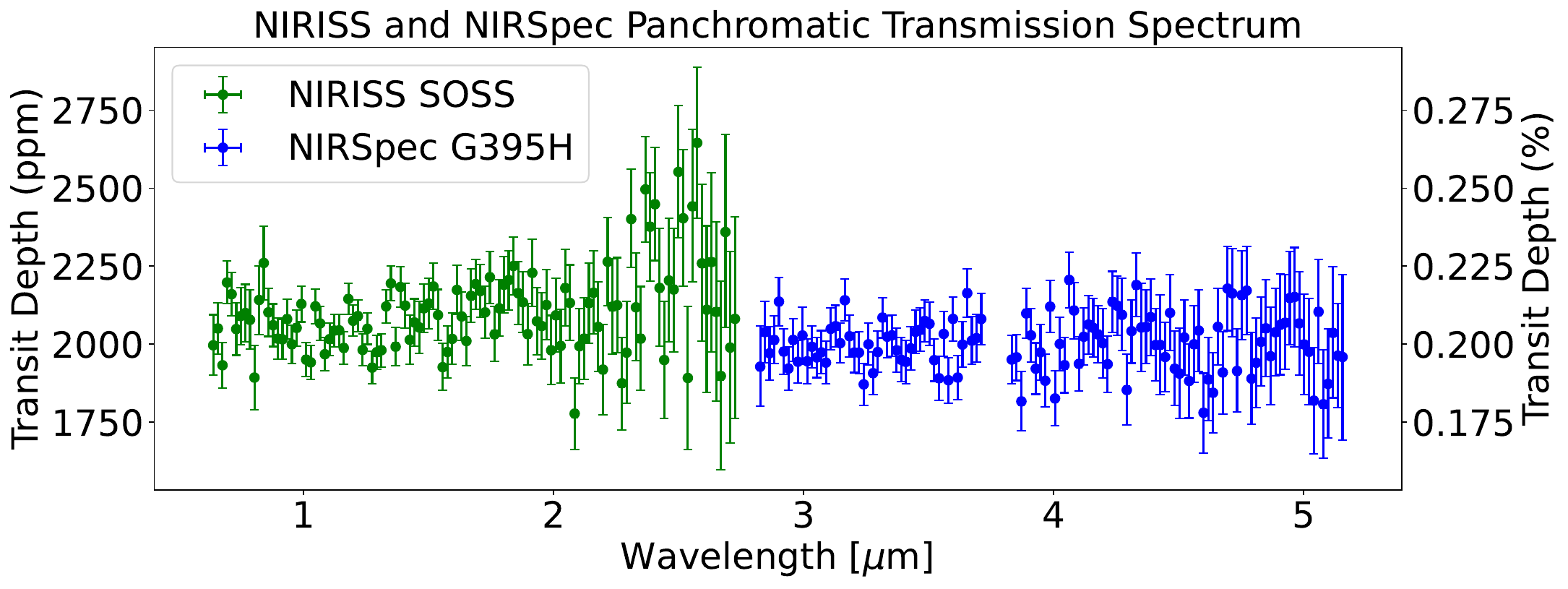}

    \caption{Panchromatic \texttt{Eureka!} transmission spectrum of LTT 9779 b combining JWST/NIRISS SOSS and JWST/NIRSpec G395H observations over 0.63-5.165~$\mu$m. The green points show the NIRISS/SOSS spectrum and the blue points show the NIRSpec/G395H spectrum. Transit depth is shown in ppm on the left y-axis and percent on the right y-axis.}
    \label{fig:spectrum}
\end{figure*}

\begin{table*}[t!]
    \centering
    \captionsetup{width=0.95\textwidth}

    \caption{
    White-light-curve fitting parameters and channel-weighted spectroscopic 
    $R_p/R_*$ values. All values that are not fixed from 
    \citealt{2025A&A...700A..45S} are rounded to 4 decimal places for readability. 
    BMJD = BJD - 2400000.5. The reported channel-weighted $R_p/R_*$ values are 
    inverse-variance weighted means of the retained spectroscopic channels and 
    are not broadband white-light-fit parameters.
    }
    \label{tab:lc_fit}

    \vspace{0.5em}

    \resizebox{0.98\textwidth}{!}{%
    \begin{tabular}{
        l
        c
        c
        c
        c
        c
    }
        \hline \hline
        Parameter 
        & Prior 
        & NRS1 
        & NRS2 
        & SOSS Order 1 
        & SOSS Order 2 \\ 
        \hline
        
        \multicolumn{6}{l}{\textbf{White-Light Planet and Orbit Parameters}} \\
        
        $R_p/R_*$ 
        & $\mathcal{U}(0.03, 0.06)$ 
        & -- 
        & -- 
        & -- 
        & -- \\

        $P_{\rm orb}$ (d) 
        & 0.792064182 
        & Fixed 
        & Fixed 
        & Fixed 
        & Fixed \\

        $i$ (\degree{}) 
        & $\mathcal{U}(70, 80)$
        & Joint 
        & Joint 
        & Joint 
        & Joint \\

        $a/R_*$  
        & $\mathcal{U}(3, 5)$
        & Joint 
        & Joint 
        & Joint 
        & Joint \\

        $e$  
        & 0.0 
        & Fixed 
        & Fixed 
        & Fixed 
        & Fixed \\

        $\omega$ (\degree{}) 
        & 90 
        & Fixed 
        & Fixed 
        & Fixed 
        & Fixed \\

        $t_{0,\mathrm{NIRSpec}}$ (BMJD) 
        & $\mathcal{U}(60600.78, 60600.82)$ 
        & Joint 
        & Joint 
        & -- 
        & -- \\ 

        $t_{0,\mathrm{NIRISS}}$ (BMJD) 
        & $\mathcal{U}(59767.4, 59767.7)$
        & -- 
        & -- 
        & Joint 
        & Joint \\
        \hline

        \multicolumn{6}{l}{\textbf{White-Light Systematics Parameters}} \\

        $c_0$ 
        & $\mathcal{N}(1, 0.05)$ 
        & $1.0004^{+4.6455\times10^{-5}}_{-4.6713\times10^{-5}}$ 
        & $1.0004^{+2.0323\times10^{-5}}_{-2.1509\times10^{-5}}$ 
        & $1.0002^{+2.0383\times10^{-5}}_{-2.0970\times10^{-5}}$ 
        & $1.0002^{+2.3929\times10^{-5}}_{-2.3657\times10^{-5}}$ \\

        $c_1$ 
        & $\mathcal{N}(0, 0.01)$ 
        & $-0.0011^{+0.0011}_{-0.0011}$ 
        & $-5.6311\times10^{-6} \pm 0.0005$ 
        & $-0.0019^{+0.0003}_{-0.0003}$ 
        & $-0.0021^{+0.0004}_{-0.0004}$ \\
        \hline

        \multicolumn{6}{l}{\textbf{Channel-Weighted Spectroscopic Values}} \\

        Weighted mean $R_p/R_*$ 
        & -- 
        & $0.044726 \pm 0.000115$
        & $0.044817 \pm 0.000151$
        & $0.045507 \pm 0.000089$
        & $0.045639 \pm 0.000275$ \\

        Weighted mean depth (ppm)
        & --
        & $2000.4 \pm 10.2$
        & $2008.6 \pm 13.5$
        & $2070.9 \pm 8.1$
        & $2082.9 \pm 25.1$ \\

        \hline
    \end{tabular}
    }
\end{table*}

\section{Independent Spectral Extractions}\label{sec:independent_extractions}
To verify the NIRSpec transmission spectrum, we utilized independent spectral extractions (Figure \ref{fig:NIRSpec_spectrum_independent}) with a lower resolution to verify the robustness of the data reduction process. 

\subsection{Independent Spectral Extraction: \texttt{Tiberius}}
We independently reduce the data using the \texttt{Tiberius} pipeline \citep{kirk2017,kirk2021}\footnote{\url{https://tiberius.readthedocs.io/en/latest/}}; the steps detailed here follow a similar process to that used in \citet{meech2025,2025MNRAS.537.3027K}.
We begin by processing the uncalibrated files (\texttt{uncal.fits}) using the \texttt{jwst} pipeline\footnote{\url{https://jwst-pipeline.readthedocs.io/en/latest/}} \citep{jwstpipeline}, running the group scale, dq init, saturation, superbias, linearity and dark current steps.
To correct for the background in the detector images, we implement our custom $1/f$ correction, similarly to \citet{meech2025}.
We then implement the ramp fitting and gain scale corrections steps from the \texttt{jwst} stage~1 pipeline.

Having processed the NRS1 and NRS2 detector images, we proceeded to spectral extraction. 
We identify the central peak of the trace by fitting each column with a Gaussian function, along the dispersion axis. 
We then sum the counts in each spectral column, over a defined aperture width, centered on the fitted trace.
The result is a time-series of stellar spectra.
We resampled the spectra: using the first spectrum in the time-series as a reference, we cross-correlated all spectra against it and apply a global shift in the dispersion direction, according to the peak correlation.
For the wavelength solution, we utilise the solution afforded by the \texttt{assign\_wcs} step from the \texttt{jwst} pipeline. 

\begin{figure*}[t]
    \centering{}
    \includegraphics[width=1.0\linewidth, trim={0cm 0cm 0cm 0cm}, clip]{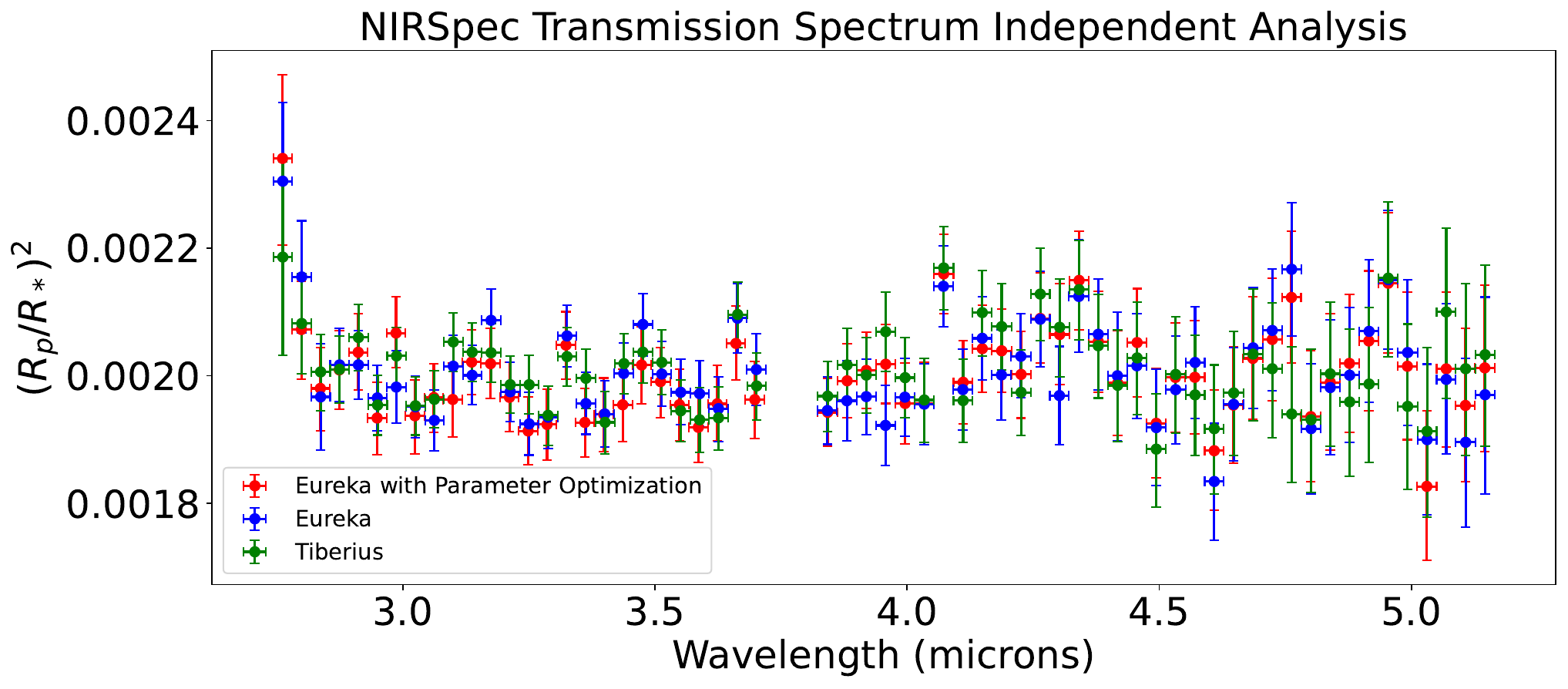}
    \caption{Comparison of three independent reductions of the JWST/NIRSpec G395H transit data. The green points show the primary \texttt{Eureka!} reduction, the red points show the parameter-optimized \texttt{Eureka!} reduction, and the blue points show the independent \texttt{Tiberius} reduction. The agreement among the three spectra demonstrates that the overall transmission-spectrum shape is robust to the choice of extraction and light-curve analysis procedure.}
    \label{fig:NIRSpec_spectrum_independent}
\end{figure*}

The task then is to construct the transit light curves: we construct both a broadband white light curve, by summing the counts across the entire dispersion axis, as well as a series of spectroscopic light curves, summing the counts in pre-defined wavelength bins.
We fit these light curves with a combined \texttt{batman}\footnote{\url{https://lkreidberg.github.io/batman/docs/html/index.html}} \citep{Kreidberg2015} transit $+$ systematics model, finding a simple linear-in-time polynomial to afford sufficient detrending for the latter.
We apply sigma-clipping to the light curves prior to fitting to remove outliers, trimming any points that diverge further than $4\sigma$ from the median.
For the fitting process, we use the \texttt{emcee}\footnote{\url{https://emcee.readthedocs.io/en/stable/}} MCMC sampler to sample the posteriors, instigating 10,000 walkers \citep{Foreman-Mackey2013}.
First we fit the white light curves (of each detector separately), leaving the mid-transit time, $a/R_*$, $i$, $R_\mathrm{p}/R_*$ and the parameters of the systematics model as free parameters.
For all light curves, we fix the limb-darkening coefficients to the solution given by the \texttt{ExoTiC-LD} package, using a quadratic limb-darkening law and the 3D \texttt{stagger} stellar model grid solution.
For the spectroscopic light curve fitting, we fix the mid-transit time, $a/R_*$, and $i$ to the corresponding detector white light curve value, and fit only for $R_\mathrm{p}/R_*$ and the parameters of the systematics model.
The resulting transmission spectrum is shown in Fig.~\ref{fig:NIRSpec_spectrum_independent}.

\subsection{Independent Spectral Extraction: \texttt{Eureka!} with Parameter Optimization}
We began the spectroscopic light-curve analysis using the Stage 3 outputs from the parameter-optimized \texttt{Eureka!} reduction. The data were reduced following the general process described in \cite{2025AJ....169..106A}, with the same reduction parameters adopted for the NRS1 and NRS2 detectors as shown in Table \ref{tab:lc_fit}. In addition, we included an exponential ramp model in the NRS1 light-curve fitting to account for instrument systematics. To isolate the transit, we manually clipped the integrations for both spectral orders. The resulting white-light curves were constructed over wavelength ranges of 2.743-3.717 $\mu$m for NRS1 and 3.823-5.165 $\mu$m for NRS2. During the fitting process, stellar limb-darkening coefficients were calculated using the mps2 stellar atmosphere grid \citep{2022A&A...666A..60K}. This reduction and fitting approach follows the methodology applied to the NIRSpec phase curve analysis of LTT 9779 b in \cite{2026AJ....171..215A}.

\section{Atmospheric Retrievals and Cloud Modeling}
We used \texttt{petitRADTRANS} (pRT; \citealt{2019A&A...627A..67M, 2024JOSS....9.7028B}) for atmospheric retrievals \citep{2024JOSS....9.5875N} to explore what atmospheric scenarios are consistent with the transmission spectrum data. Additionally, we utilized an independent equilibrium chemistry atmospheric retrieval (Figure \ref{fig:eqm_independent_analysis}) on the panchromatic spectrum to verify the retrieval process. To further explore the cloud parameters, we used forward model grids combining \texttt{picaso} \citep{2019ApJ...878...70B} and \texttt{virga} \citep{2025ApJ...994..116M, 2026AJ....171...98B}.

\subsection{Atmospheric Retrievals}\label{subsec:retrievals}
We started with pRT free-chemistry retrievals that include the following species: \ce{H2O} \citep{2010JQSRT.111.2139R}, \ce{CO} \citep{2010JQSRT.111.2139R}, \ce{CO2} \citep{2020MNRAS.496.5282Y}, \ce{SO2} \citep{2016MNRAS.459.3890U}, \ce{CH4} \citep{2020ApJS..247...55H}, and \ce{OCS} \citep{2024MNRAS.530.4004O}. Most of these species are expected to be present in exoplanet atmospheres and were therefore included in the retrievals. We additionally included \ce{OCS} because it can become abundant at high metallicities and may contribute to the feature in transmission near $\sim$4.7~$\mu$m \citep{2023Natur.614..664A, 2025MNRAS.537.3027K}. Since LTT 9779 b is expected to have an \ce{H2}/\ce{He}-dominated atmosphere \citep{2020ApJ...903L...7C, 2020ApJ...903L...6D}, our Rayleigh-scattering species are \ce{H2} and \ce{He}, with collision-induced absorption from \ce{H2}-\ce{H2} \citep{2001JQSRT..68..235B, 2002A&A...390..779B} and \ce{H2}-\ce{He} \citep{1988ApJ...326..509B, 1989ApJ...336..495B, 1989ApJ...341..549B}. We use a reference pressure of 0.01 bar \citep{2024ApJ...961L..23B} and run the retrievals over a log pressure range of $(-8,3)$. We assume an isothermal temperature structure, with the atmospheric temperature treated as a free parameter. We also retrieve an offset between the NIRISS and NIRSpec transmission spectra to account for differences in their absolute transit-depth levels. We first performed a retrieval using these parameters to obtain a cloud-free atmospheric model. To incorporate clouds into the free-chemistry retrievals, we then utilized a patchy opaque cloud deck, fitting for the cloud-top pressure and the cloud fraction. All priors used in the retrievals are shown in Table \ref{tab:retrieval_priors}.

\begin{table}[h]
    \centering
    \begin{tabular}{lc} \hline \hline
        Parameter & Prior \\ \hline
        Planet Radius ($R_{\oplus}$) & $\mathcal{U}(3, 6)$ \\
        NIRISS/NIRSpec Offset & $\mathcal{U}(-5 \times 10^{-4}, 5 \times 10^{-4})$ \\ \hline
        Temperature (K) & $\mathcal{U}(0,3000)$ \\
        \ce{H2O} & $\mathcal{U}(-10, 0)$ \\
        \ce{CO} & $\mathcal{U}(-10, 0)$  \\
        \ce{CO2} & $\mathcal{U}(-10, 0)$ \\
        \ce{SO2} & $\mathcal{U}(-10, 0)$ \\
        \ce{CH4} & $\mathcal{U}(-10, 0)$ \\ 
        \ce{OCS} & $\mathcal{U}(-10, 0)$ \\ \hline
        Temperature (K) & $\mathcal{U}(500,3000)$ \\
        $\log_{10}(\text{Metallicity})$ & $\mathcal{U}(-1, 3)$\\
        C/O & $\mathcal{U}(0, 1.6)$ \\ \hline
       $\log_{10}(\text{Opaque Cloud-}$ \\ \hspace{0.5em}\text{Top Pressure}) & $\mathcal{U}(-8, 0)$ \\
        Cloud Fraction & $\mathcal{U}(0, 1)$ \\ \hline
    \end{tabular}
    \caption{Prior ranges of our full free chemistry and equilibrium chemistry retrievals. $\mathcal{U}(a,b)$ denotes a uniform distribution bounded by $a$ and $b$. The first set of parameters is used in all retrievals, the second set of parameters is used in free chemistry retrievals, the third set of parameters is used in equilibrium chemistry retrievals, and the fourth set of parameters is used in cloudy, free and equilibrium retrievals (for a retrieval with a 100\% cloud fraction, the Cloud Fraction parameter is set to 1).}
    \label{tab:retrieval_priors}
\end{table}

We implement a leave-one-out process \citep[e.g.,][]{2009ApJ...690.1056M, 2013ApJ...778..153B}, in which we exclude one molecule from the retrieval at a time, to determine if any of the species are significantly detected in transmission. Results from this process are shown in Section \ref{subsec:leaveoneout}. 

We then performed equilibrium chemistry retrievals using pRT's pre-calculated equilibrium chemistry table. We used the following line species to calculate the atmospheric composition: \ce{H2O} \citep{2010JQSRT.111.2139R}, \ce{CO} \citep{2010JQSRT.111.2139R}, \ce{CO2} \citep{2020MNRAS.496.5282Y}, \ce{FeH} \citep{2010A&A...523A..58W}, \ce{SiO} \citep{2022MNRAS.510..903Y}, \ce{TiO} \citep{2019MNRAS.488.2836M}, \ce{VO} \citep{2016MNRAS.463..771M}, \ce{HCN} \citep{2014MNRAS.437.1828B}, \ce{K} \citep{2019A&A...627A..67M}, and \ce{Na} \citep{2019A&A...628A.120A}. The equilibrium-chemistry retrievals include additional opacity sources compared to the free-chemistry retrievals because the number of species included in the latter was limited to avoid introducing a large number of additional free abundance parameters. In the equilibrium-chemistry framework, however, the abundances of these species are determined self-consistently by the retrieved atmospheric properties rather than being independently fit. We therefore include additional species, particularly high-temperature opacity sources such as \ce{FeH}, \ce{SiO}, and \ce{TiO}, whose presence or absence can provide information about the equilibrium atmospheric composition without introducing additional free abundance parameters. \ce{CH4} and sulfur-bearing species were not included as line-opacity sources in these retrievals because \ce{CH4} is expected to be relatively unimportant at the high temperatures probed here, while several sulfur-bearing species considered in the free-chemistry retrievals are not included in the standard pre-calculated equilibrium chemistry abundance table. Our Rayleigh species are \ce{H2} and \ce{He}, with collision-induced absorption from \ce{H2}-\ce{H2} and \ce{H2}-\ce{He}. We do not impose a set ratio of \ce{H2} to \ce{He} within the atmosphere, instead allowing the calculated equilibrium-chemistry abundances of these species to be used. We use a reference pressure of 0.01 bar \citep{2024ApJ...961L..23B} and run the retrievals over a log pressure range of $(-8,3)$, as in the free-chemistry fits. We perform two cloudy retrievals: one with a patchy cloud deck and one with a 100\% cloud fraction. All priors used in the retrievals are shown in Table \ref{tab:retrieval_priors}.

All retrievals used \texttt{MultiNest} \citep{2008MNRAS.384..449F, 2009MNRAS.398.1601F, 2019OJAp....2E..10F} within pRT, with 1,000 live points and a sampling efficiency of 0.3. In the free chemistry retrievals, we retrieve the atmosphere's temperature, the planetary radius at the reference pressure, and the $\mathrm{log}_{10}$(MMR) for each species included in the retrieval. In the equilibrium chemistry retrievals, instead of retrieving the individual species MMR, we retrieve the atmospheric metallicity and C/O ratio. We retrieve the opaque cloud-top pressure and cloud fraction for the patchy cloud retrievals, and the cloud-top pressure for the 100\% cloud fraction retrievals. The best-fit spectra from the patchy-cloud free-chemistry, patchy-cloud equilibrium-chemistry, and 100\% cloud-fraction equilibrium-chemistry retrievals are compared in Figure \ref{fig:retrieval_comparison}. The full posterior distributions for the patchy-cloud free-chemistry and patchy-cloud equilibrium-chemistry retrievals are shown in Section \ref{subsec:cloudy_retrieval_posteriors}.

\subsection{Independent Atmospheric Retrieval}

Also using pRT, we performed an independent retrieval of the data to verify our results. In these retrievals, we used pRT's built-in models (\texttt{guillot\_transmission} and \texttt{isothermal\_transmission}) rather than building the temperature structure, mass fractions, and resulting spectra manually. In this setup, we assumed chemical equilibrium, fitting for metallicity and C/O with the same species and lines as above (i.e., H$_2$O, CO, CO$_2$, FeH, SiO, TiO, VO, HCN, K, and Na). A cloud top pressure was also retrieved, but assuming a 100\% cloud fraction, rather than allowing for patchy clouds. We tested both a one-parameter isothermal atmosphere and the three-parameter profile from \cite{2010A&A...520A..27G}. Lastly, we allowed for instrumental offsets between the two NIRISS orders, the two NIRSpec detectors, and the two instruments themselves.

A lower Bayesian evidence for the three-parameter temperature profile ($\ln B = -0.9$) indicates that the extra parameters of the non-isothermal profile were not justified by the data. We therefore describe and compare below the results of the isothermal retrieval setup.

During comparison of the independently implemented equilibrium-chemistry retrievals, we identified a difference in the treatment of the background \ce{H2}/\ce{He} atmosphere. In an earlier implementation, \ce{H2} and \ce{He} were forced to fill the remaining atmospheric mass fraction in a fixed ratio after the equilibrium-chemistry abundances were calculated. At high metallicity, this treatment suppressed the expected increase in atmospheric mean molecular weight and consequently altered the atmospheric scale height and spectral-feature amplitudes, disfavoring the high-metallicity mode recovered by the independent retrieval. We therefore adopt the equilibrium-chemistry retrieval in which the abundances of \ce{H}, \ce{H2}, and \ce{He} are determined self-consistently by the equilibrium-chemistry calculation, without imposing an additional filling ratio.

\begin{figure*}[!htb]
    \centering{}
    \includegraphics[width=1.0\linewidth, trim={0cm 0cm 0cm 0cm}, clip]{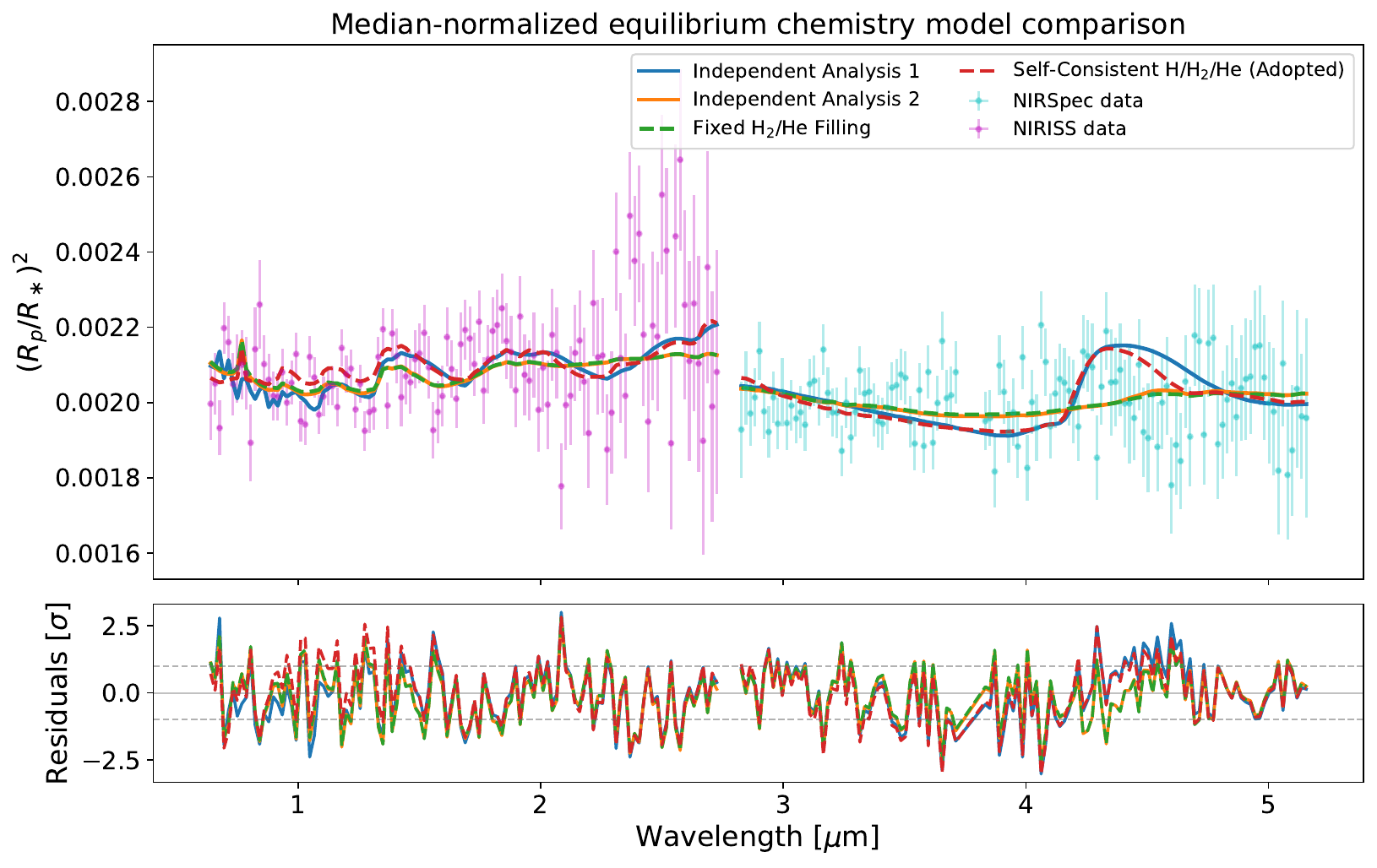}
    \caption{Comparison of four 100\% cloud-fraction equilibrium-chemistry retrieval setups for the panchromatic NIRISS/NIRSpec transmission spectrum. The red curve shows our adopted self-consistent treatment, in which the \ce{H}, \ce{H2}, and \ce{He} abundances are set by the equilibrium-chemistry calculation, while the green curve shows an earlier implementation that forced \ce{H2} and \ce{He} to fill the remaining atmospheric mass fraction in a fixed ratio. The blue and orange curves show two independent retrievals used to validate the retrieval framework. The imposed filling treatment alters the high-metallicity behavior through its effect on the atmospheric mean molecular weight and scale height. The lower panel shows residuals in units of the observational uncertainty.}
    \label{fig:eqm_independent_analysis}
\end{figure*}

Figure \ref{fig:eqm_independent_analysis} shows a comparison of the adopted 100\% cloud-fraction equilibrium-chemistry retrieval described in Section \ref{subsec:retrievals}, a second 100\% cloud-fraction retrieval in which \ce{H2} and \ce{He} are forced to fill the remaining atmospheric mass fraction in a fixed ratio, and two independent retrievals. The comparison with the independent retrievals helped identify the effect of the imposed \ce{H2}/\ce{He} filling prescription at high metallicity and motivated our adoption of the no-imposed-filling treatment. With this treatment, the primary and independent retrieval frameworks show consistent behavior, including recovery of the high-metallicity posterior mode. A more detailed comparison of the imposed- and no-imposed-filling retrievals is shown in Section \ref{subsec:independent_retrieval_zoomin}. 

\subsection{Cloud Forward Modeling}\label{subsec:models}
We run two forward model grids and examine their results independently and jointly to further explore the cloud properties of LTT 9779 b.

The first grid uses \texttt{virga} and \texttt{picaso} self-consistent models based on \cite{2001ApJ...556..872A} to examine the impact of different $f_{\rm sed}$ (sedimentation efficiency, or how effectively cloud condensate particles settle downward under gravity compared to how fast they are mixed upward by atmospheric turbulence) values, $K_{zz}$ values, and widths of the particle size distribution on the modeled spectrum. The condensate used in the main grid is \ce{Mg2SiO4}, which was suggested as a possible cloud species for LTT 9779 b based on the planet's albedo \citep{2025NatAs...9..512C}. The underlying gas atmosphere was fixed to the best-fit retrieved free-chemistry cloudy transmission abundances and an isothermal temperature profile of 561 K, while the \texttt{virga} cloud calculation used a metallicity of $10^{1.84}$ (from the 100\% cloud fraction equilibrium-chemistry retrieval) and a fixed mean molecular weight of 2.3. The $f_{\rm sed}$ values in the grid are [0.1, 0.3, 1.0, 3.0, 6.0]. The $k_{zz}$ values in the grid are $[10^8, 10^9, 10^{10}, 10^{11}] \ \mathrm{cm}^2 \ \mathrm{s}^{-1}$. For each \texttt{virga} model, the atmosphere DataFrame (analogous to a \texttt{pandas} DataFrame, \citealt{mckinney-proc-scipy-2010}) includes a constant $k_{zz}$ profile equal to the gridded value for that model. Thus, $k_{zz}$ is varied across the grid but is vertically constant within each individual model. We also grid over the width of the particle size distribution with values [1.1, 1.5, 2.0, 2.5, 3.0]. The overall grid has 100 models in total. 

The second grid uses a custom cloud slab in \texttt{virga} and \texttt{picaso} to examine the effects of the mean cloud particle radius, the cloud's base and top pressure, and the particle column density. Unlike the first grid, this grid does not include $f_{\rm sed}$, since the slab structure is imposed directly rather than calculated from a sedimentation model. Although the custom slab cloud structure is imposed directly, we still include a fixed $k_{zz}$ = $10^9$ $\mathrm{cm}^2 \mathrm{s}^{-1}$ column in the atmosphere profile because \texttt{picaso} expects this field for cloudy atmosphere calculations. We use the same condensate as the first grid. For the Gaussian particle-size distribution, we use $\sigma_r = 0.5$ in $\log_{10}(r)$ space. For the grid of the mean particle radius, we use $\log_{10}(r)$ values of [-6.5, -6.0, -5.5, -5.0, -4.5, -4.0, -3.5, -3.0], where $r$ is in cm. These correspond to mean radii of approximately 0.0032, 0.01, 0.032, 0.1, 0.32, 1.0, 3.2, and 10 $\mu$m, respectively. We use a cloud-base pressure grid of [3.0, 1.0, 0.3, 0.1] bars. For the cloud-top pressure, we grid over [$10^{-2}$, $10^{-3}$, $10^{-4}$] bars and include the best-fit cloud-top pressure from the equilibrium-chemistry retrieval, $10^{-1.5}$ bar, as an additional grid point. Our column density grid is [$10^4$, $10^5$, $10^6$, $10^7$] particles/$\mathrm{cm}^2$. The overall grid has 512 models in total.

For both grids, each model spectrum is binned to the resolution of the extracted NIRISS and NIRSpec spectra (see Sections \ref{subsec:NIRISS_reduction} and \ref{subsec:NIRSpec_reduction}) before comparison with the data. We then fit a wavelength-independent vertical offset between the model and observed transit depths and calculate both $\chi^2$ and reduced $\chi^2$. The models from both grids are ranked separately by reduced $\chi^2$ to identify the best-fitting cloud configurations. 

In addition, we run a fixed-$f_{\rm sed}$ condensate comparison with $f_{\rm sed}=1$ using the best-fit \texttt{virga} parameters from the first grid. This comparison includes \ce{Mg2SiO4}, \ce{MgSiO3}, \ce{Fe}, \ce{Al2O3}, and \ce{TiO2}, which were suggested cloud species from \texttt{virga} for a solar-metallicity atmosphere with high temperature and mean molecular weight consistent with a \ce{H}/\ce{He}-dominated atmosphere \citep{2025NatAs...9..512C}. We test whether additional condensate species produce substantially different spectral behavior. Section \ref{subsec:cloud_condensates} shows the spectral comparison for the cloud species.

\section{Results and Discussion}\label{sec:result_discuss}

\begin{figure*}[!htb]
    \centering
    \includegraphics[
        width=1.0\linewidth,
        trim={0cm 0cm 0cm 0cm},
        clip
    ]{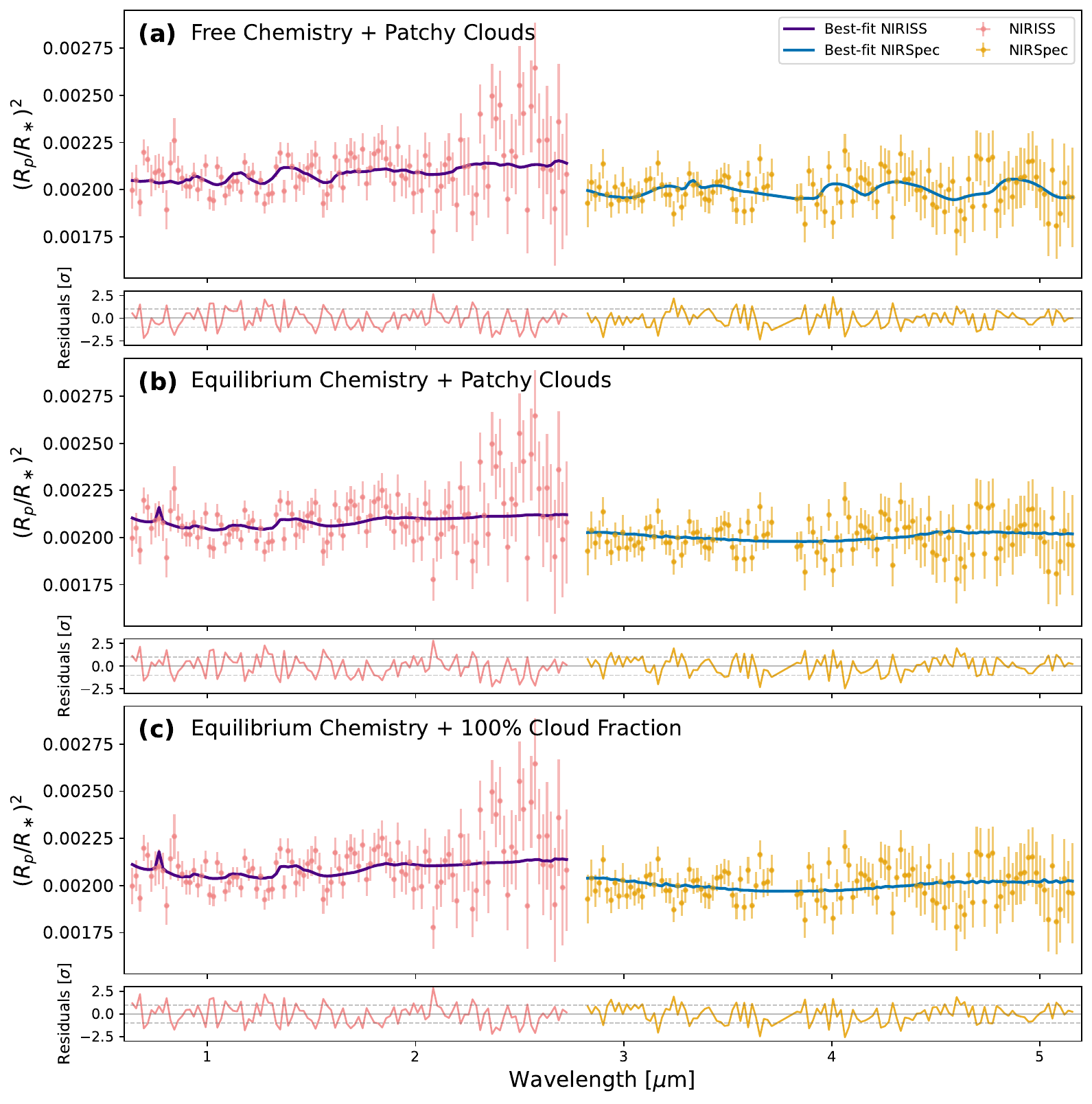}
    \caption{
    Comparison of best-fit atmospheric retrievals for the panchromatic NIRISS/NIRSpec transmission spectrum. Panel (a) shows the patchy-cloud free-chemistry retrieval, panel (b) shows the patchy-cloud equilibrium-chemistry retrieval, and panel (c) shows the 100\% cloud-fraction equilibrium-chemistry retrieval. In each panel, the upper axis shows the observed NIRISS and NIRSpec transit depths with the corresponding best-fit model spectra, while the lower axis shows residuals in units of the observational uncertainty. The three retrieval frameworks illustrate how different assumptions about atmospheric chemistry and cloud coverage can reproduce the muted transmission spectrum.
    }
    \label{fig:retrieval_comparison}
\end{figure*}

\begin{figure*}[!htb]
    \centering
    \includegraphics[
        width=1.0\linewidth,
        trim={0cm 0cm 0cm 0cm},
        clip
    ]{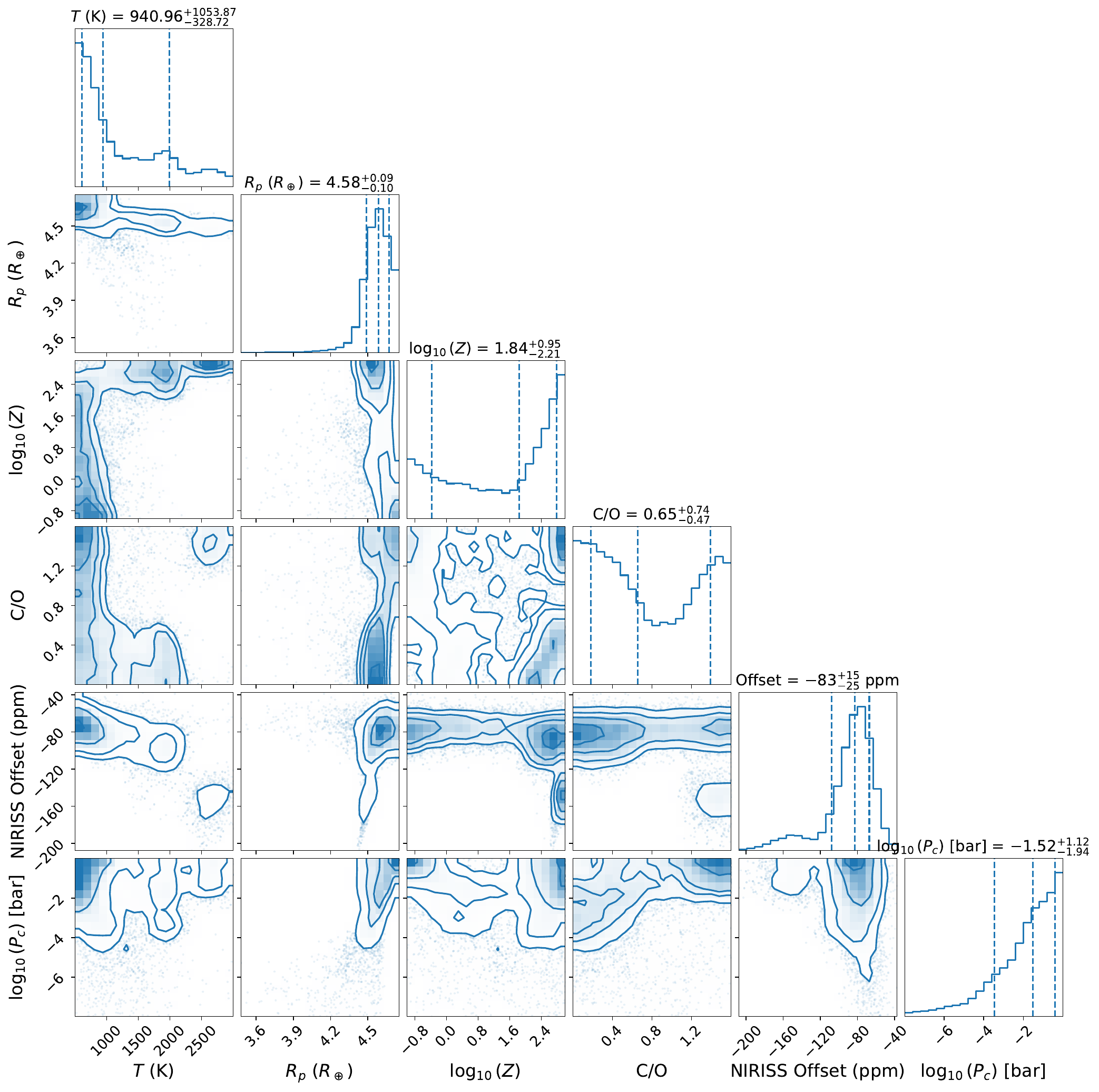}
    \caption{
    Posterior distributions for the 100\% cloud-fraction equilibrium-chemistry retrieval. Retrieved parameters include the isothermal atmospheric temperature, planetary radius, atmospheric metallicity, C/O ratio, NIRISS/NIRSpec offset, and opaque cloud-top pressure.
    }
    \label{fig:eqm_100clouds_cornerplot}
\end{figure*}

\begin{figure*}[!htb]
    \centering{}
    \includegraphics[width=1.0\linewidth, trim={0cm 0cm 0cm 0cm}, clip]{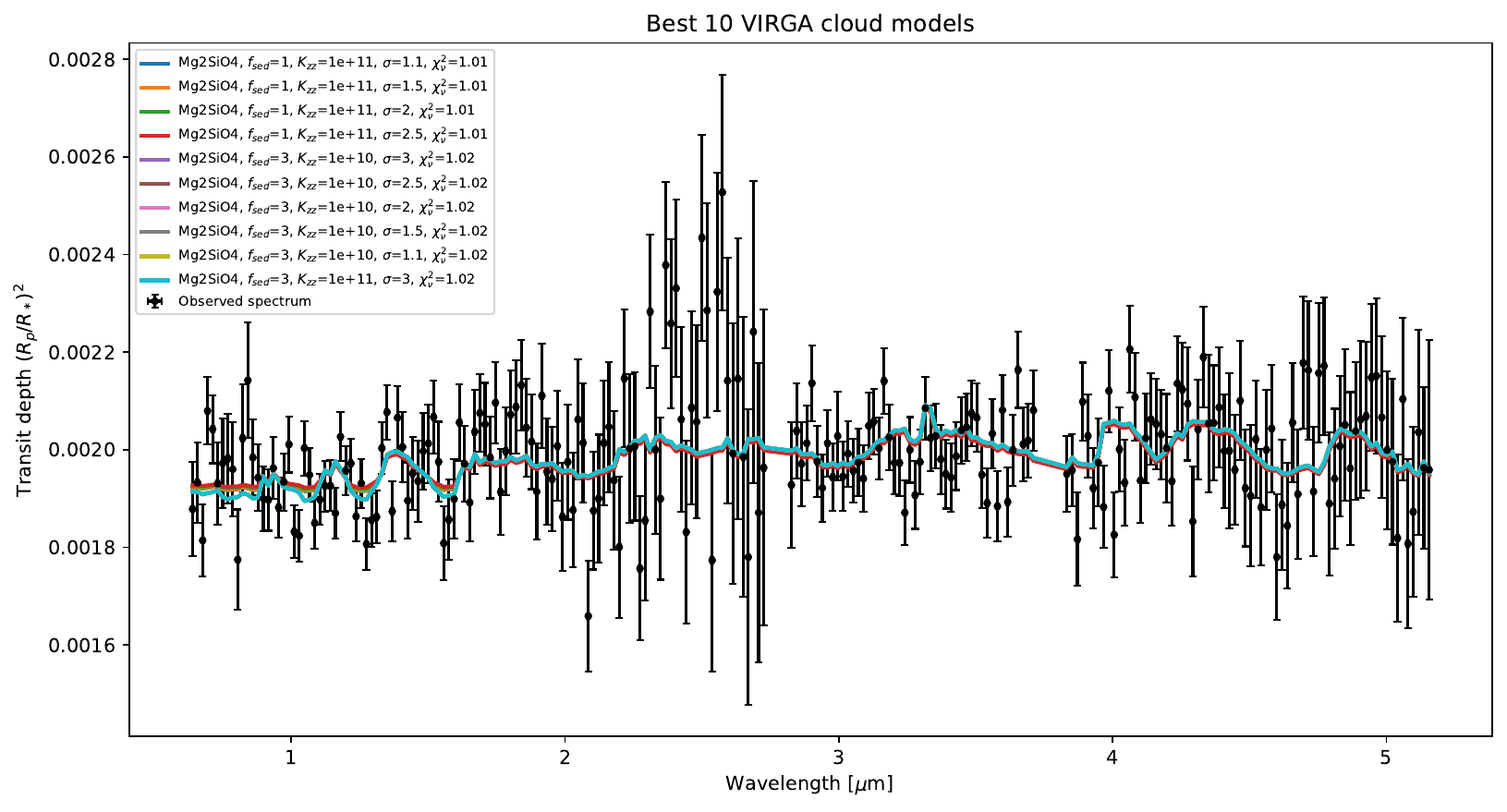}
    \caption{Ten best-fitting \texttt{virga}+\texttt{picaso} \ce{Mg2SiO4} cloud forward models compared with the observed panchromatic transmission spectrum. The models span sedimentation efficiency $f_{\rm sed}$, eddy diffusion coefficient $K_{zz}$, and particle-size distribution width $\sigma$. The legend gives the cloud parameters and reduced $\chi^2$ for each model.}
    \label{fig:virga}
\end{figure*}

\begin{figure*}[!htb]
    \centering{}
    \includegraphics[width=1.0\linewidth, trim={0cm 0cm 0cm 0cm}, clip]{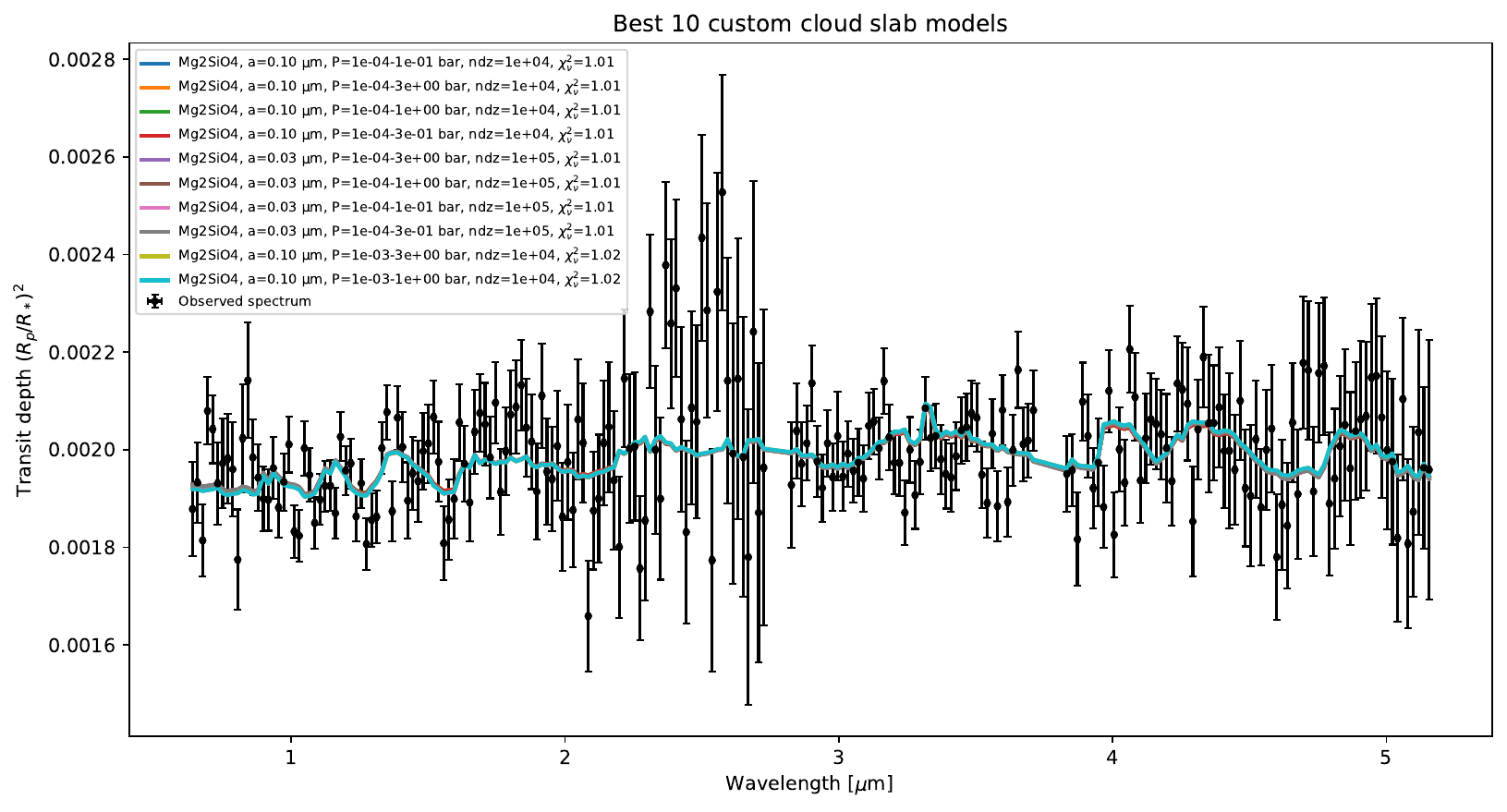}
    \caption{Ten best-fitting custom cloud-slab \ce{Mg2SiO4} forward models compared with the observed panchromatic transmission spectrum. The models vary the mean particle radius, cloud-top and cloud-base pressures, and particle column density. The legend lists the corresponding cloud parameters and reduced $\chi^2$ values.}
    \label{fig:customslab}
\end{figure*}

\subsection{Cloudy Nights}

The relatively muted transmission spectrum of LTT 9779 b naturally motivates the presence of clouds at the terminator, particularly given the multiple lines of evidence for condensates elsewhere in the planet's atmosphere. Previous studies have inferred silicate clouds from the planet's high optical albedo, reflected-light measurements, and NIRISS transmission spectrum, with cloud opacity suggested at approximately millibar pressures \citep{2023A&A...675A..81H,radica2024muted,2025A&A...700A..45S,2025NatAs...9..512C,2026ApJ..1005L..79S}. Our patchy-cloud free-chemistry retrieval provides a good fit to the muted transmission spectrum (Figure \ref{fig:retrieval_comparison}a), although our transmission data do not statistically require clouds. The cloudy and clear free-chemistry retrievals have comparable Bayesian evidence, with $\ln Z=-130.5$ and $-129.7$, respectively (Table \ref{tab:significances}), corresponding to only a weak preference for the clear model. Moreover, the molecular abundances inferred from the clear transmission retrieval are broadly consistent with those independently obtained from clear-atmosphere emission retrievals \citep{Brande2026} for \ce{H2O}, \ce{CO2}, and \ce{CO}. Thus, a clear terminator cannot be excluded from the transmission spectrum alone. Nevertheless, clouds remain a physically well-motivated explanation for the weak spectral modulation, particularly because transmission spectroscopy is especially sensitive to aerosol opacity, while independent observations indicate substantial cloud coverage elsewhere on the planet.

Our 100\% cloud fraction equilibrium-chemistry retrieval (Figure \ref{fig:retrieval_comparison}c) favors a high-metallicity branch of the posterior ($\sim70 \times$ solar), although sub-solar metallicities remain allowed within 1$\sigma$. The corresponding posterior distributions are shown in Figure \ref{fig:eqm_100clouds_cornerplot}. The broad metallicity constraint is strongly correlated with the retrieved atmospheric temperature. In particular, the posterior extends toward temperatures substantially below the expected equilibrium temperature of LTT 9779 b. Given the muted transmission spectrum, these low-temperature solutions likely reflect a retrieval degeneracy rather than a physically cold terminator. Because the atmospheric scale height scales approximately as $H \propto T/\mu g$, decreasing the temperature reduces the amplitude of spectral features in much the same way as increasing the mean molecular weight through higher metallicity. Consequently, at lower retrieved temperatures, a wide range of metallicities, including sub-solar values, can reproduce the relatively flat spectrum, whereas at higher temperatures the posterior increasingly favors higher metallicity.

Our results are therefore consistent with previously reported evidence for a high-metallicity atmosphere while demonstrating that the current panchromatic transmission spectrum alone does not uniquely require such high metallicities. The extent to which the retrieval exploits very low temperatures to reproduce the muted spectrum also suggests that the inferred metallicity and cloud properties may be sensitive to the adopted temperature prior. Future retrievals employing more physically constrained temperature priors will be useful for determining whether restricting these low-temperature solutions results in stronger constraints on atmospheric metallicity or cloud opacity. This distinction is particularly important because previous metallicity constraints have been derived using a combination of different observational techniques that probe different regions of the planet's atmosphere, whereas transmission spectroscopy primarily probes the terminator. If the planet possesses strong longitudinal variations in temperature, cloud coverage, or composition, the atmospheric properties inferred from different observations may not be identical. The longitudinal cloud asymmetry already inferred from phase-curve and reflected-light observations provides a physically plausible basis for such differences.

When we instead allow the equilibrium-chemistry retrieval to vary the cloud fraction (Figure \ref{fig:retrieval_comparison}b), the atmospheric interpretation changes substantially. The retrieval shifts toward lower metallicities and favors a cloud fraction of roughly 40\%. This result illustrates the strong degeneracy between clouds and atmospheric composition in transmission. In comparison, the 100\% cloudy retrieval shown in Figure \ref{fig:retrieval_comparison}c can reproduce weak spectral features through a combination of an opaque cloud deck, a reduced atmospheric scale height, and high atmospheric metallicity. Once the terminator is allowed to probe patchy cloud fractions, however, the observed spectrum can be reproduced by a combination of cloudy and clear regions without requiring the same high metallicities. The metallicity constraint is therefore sensitive not only to whether clouds are included, but also to the distribution of those clouds across the terminator. This is especially relevant for LTT 9779 b because previous phase-curve and reflected-light measurements already suggest substantial spatial variations in cloud coverage across the planet. A heterogeneous terminator is therefore physically plausible. The current panchromatic transmission spectrum does not uniquely distinguish between a high-metallicity atmosphere with nearly complete cloud coverage and a lower-metallicity atmosphere with patchier clouds, representing one of the dominant uncertainties in determining the atmospheric composition of LTT 9779 b from transmission spectroscopy alone.

The similar quality of the fits across the three retrieval setups in Figure \ref{fig:retrieval_comparison} emphasizes that the muted transmission spectrum can be reproduced under substantially different assumptions about atmospheric chemistry and cloud coverage.

The difference between the fully cloudy and patchy-cloud retrievals also emphasizes that metallicity estimates are strongly model-dependent, especially for a muted transmission spectrum. A high-metallicity solution may suppress spectral features through the increased mean molecular weight of the atmosphere, while clouds can produce a similar effect by obscuring deeper atmospheric layers. The range of viable metallicities therefore reflects a physical degeneracy between atmospheric composition and cloud fraction.

\subsection{Cloud Properties}

\begin{figure}[ht!]
    \centering
    \includegraphics[width=1.0\columnwidth]{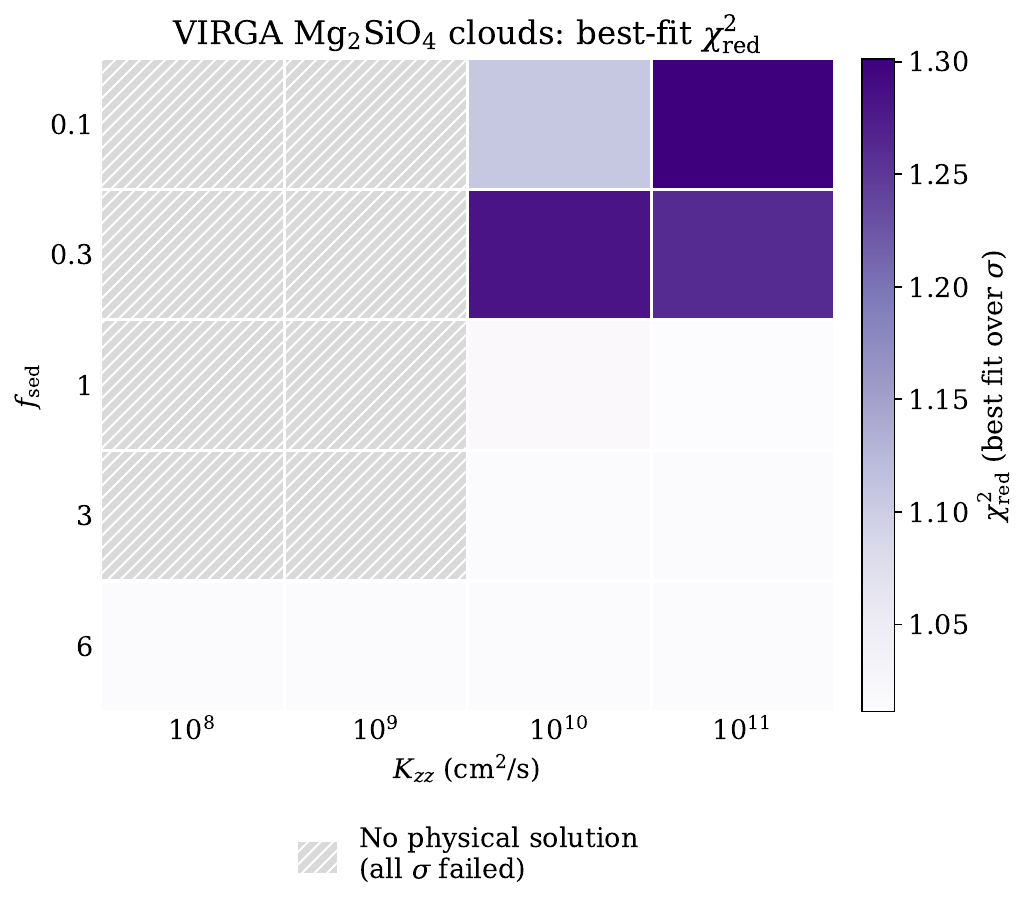}
    \caption{Best-fit reduced $\chi^2$ across the $f_{\rm sed}$-$K_{zz}$ parameter space for the \texttt{virga}+\texttt{picaso} \ce{Mg2SiO4} cloud grid. Each cell shows the minimum reduced $\chi^2$ obtained over all tested particle-size distribution widths $\sigma$ for that $(f_{\rm sed},K_{zz})$ combination. Hatched cells indicate parameter combinations for which no physical cloud solution was obtained for any value of $\sigma$.}
    \label{fig:heatmap}
\end{figure}

Having established that cloudy solutions remain physically plausible despite not being uniquely required by the retrievals, we next use our cloud forward-model grids to explore the range of cloud properties capable of reproducing the observed transmission spectrum. Within the \ce{Mg2SiO4} \texttt{virga} grid, we find a preference for moderate-to-high sedimentation-efficiency values and high $k_{zz}$ values, while the custom cloud-slab grid tends towards low particle column densities and sub-micron mean particle sizes. Figure~\ref{fig:heatmap} shows the best-fit reduced $\chi^2$ across the $f_{\rm sed}$-$k_{zz}$ grid. Models with $f_{\rm sed} \geq 1$ and $k_{zz} \geq 10^{10}$ cm$^{2}$s$^{-1}$ produce the best fits ($\chi^2_{\rm red} \approx 1.01$), while low $f_{\rm sed}$ values ($0.1$-$0.3$) yield substantially worse fits regardless of $k_{zz}$, and low $k_{zz}$ values ($10^{8}$-$10^{9}$ cm$^{2}$s$^{-1}$) fail to produce physical solutions across much of the parameter space. Taken together, these results suggest that the muted spectrum can be reproduced by a population of small cloud particles with relatively low particle column densities. The preference for moderate-to-high $f_{\rm sed}$ values in the \texttt{virga} models additionally prefers less vertically extended clouds than the low-$f_{\rm sed}$ solutions suggested by some previous albedo studies.

The preference for sub-micron particles is particularly interesting in the context of LTT 9779 b's high optical albedo. Small silicate particles can efficiently scatter incident stellar radiation at optical and near-infrared wavelengths while simultaneously muting gaseous absorption features in transmission. Our cloud modeling therefore provides a possible connection between two otherwise distinct observational properties of LTT 9779 b: its unusually reflective atmosphere and its comparatively featureless transmission spectrum may both arise from the presence of condensate clouds containing small particles. Previous albedo measurements have suggested silicate clouds with lower $f_{\rm sed}$ values, particularly $f_{\rm sed}=0.1$, implying clouds with a larger vertical extent and small particles \citep{2023A&A...675A..81H,2025A&A...700A..45S}. Our transmission modeling does not show the same preference for such low $f_{\rm sed}$ values. However, this comparison should be made cautiously because $f_{\rm sed}$ does not uniquely describe particle size and cloud extent across different cloud-modeling frameworks. In our modeling, particle size is explored separately, so we primarily interpret $f_{\rm sed}$ as a measure of how vertically extended the cloud is rather than as a parameter that simultaneously determines both particle size and cloud extent. The difference between our preferred values and those inferred from reflected-light observations may therefore partly reflect differences in model assumptions. It is also possible that the discrepancy reflects the different regions of the atmosphere probed by transmission and reflected-light observations. If LTT 9779 b possesses a spatially heterogeneous cloud distribution, as suggested by existing phase-curve observations, the cloud properties inferred from the terminator may differ from those inferred from the reflective portions of the dayside. Differences in preferred $f_{\rm sed}$ values therefore do not necessarily imply an inconsistency between the datasets.

There is no clear preference for the width of the particle-radius distribution or the pressure extent of the cloud in the custom cloud-slab grid. These parameters may have a weaker effect on the transmission spectrum over the current wavelength range or may remain degenerate with particle size, column density, and total cloud opacity. Although we adopt \ce{Mg2SiO4} as the primary condensate in our cloud grids, the transmission spectrum itself does not uniquely identify the cloud composition. The additional condensate comparison indicates that several candidate species can produce broadly similar spectral behavior over the NIRISS and NIRSpec wavelength range, limiting our ability to determine the dominant condensate species using the current data alone. It is therefore possible that \ce{Mg2SiO4} is not the primary condensate responsible for the observed cloud opacity, even if silicate clouds more generally are present. Mid-infrared observations would provide the most direct way to distinguish among these scenarios. In particular, \ce{Mg2SiO4} and \ce{MgSiO3} exhibit characteristic spectral features at longer wavelengths not sampled by the current transmission spectrum. Detecting these features would provide direct evidence for silicate condensates and help constrain their composition and particle properties. Conversely, the absence of the expected silicate signatures could indicate that other cloud species, different particle-size distributions, or more complex cloud structures are responsible for the planet's high reflectivity.

\subsection{Cool Nights}

The free-chemistry retrieval also favors a terminator temperature substantially below the planet's equilibrium temperature. The preferred temperature is roughly 560 K, compared with an equilibrium temperature of approximately 2000 K. Such a cool retrieved limb is broadly consistent with strong spatial temperature variations across the planet, given that the dayside temperatures measured from phase-curve observations are substantially higher and closer to the planet's equilibrium temperature \citep{2026AJ....171..215A}. The low preferred temperature in transmission may reflect a combination of atmospheric circulation and the low amplitude of spectral features on the cloudy terminator.

\subsection{Clouds and Chemistry}

Across all free-chemistry retrievals, we do not find evidence for statistically significant detections of individual molecular absorbers. When patchy clouds are included, however, we obtain posterior constraints on the abundances of \ce{H2O}, \ce{CO2}, \ce{SO2}, \ce{CH4}, and \ce{OCS}, as well as an upper limit on the \ce{CO} abundance. This interpretation is consistent with the Bayesian leave-one-out model comparison, which does not provide evidence for statistically significant detections of any of the included molecular species. However, the full free-chemistry model is preferred over a wavelength-independent transmission spectrum by $\Delta\ln Z \approx 3.8$, corresponding to an approximate significance of $\sim3.2\sigma$ following the conversion used by \citet{2013ApJ...778..153B}. Thus, although no individual molecule is detected with high confidence, the rejection of a flat spectrum indicates that the panchromatic transmission spectrum contains statistically significant wavelength-dependent structure. This is consistent with the combined opacity of one or more atmospheric species contributing to the observed spectral modulation, even though the present data do not uniquely identify which absorbers dominate.

The retrieval also has difficulty tightly constraining the cloud-top pressure of the opaque cloud deck. The posterior spans a broad range of pressures while allowing high-altitude clouds capable of suppressing molecular absorption. Once an opaque cloud deck is placed sufficiently high in the atmosphere, moving the cloud top to lower pressures produces little additional change in the spectrum. The spectrum is therefore more informative about the ability of clouds to mute molecular features than about their precise pressure level.

\begin{figure}[ht!]
    \centering
    \includegraphics[width=1.0\columnwidth,trim=0 170 0 170,
    clip]{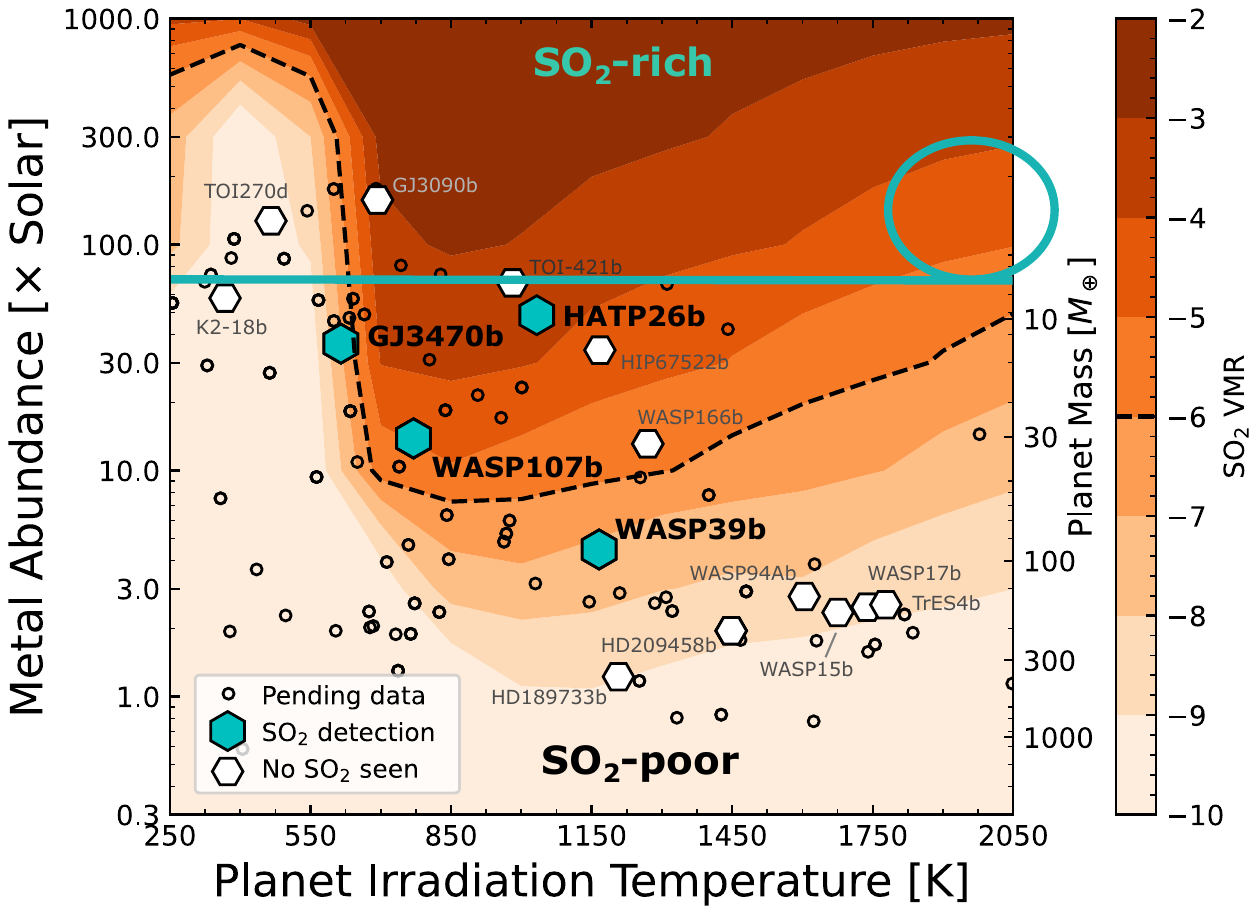}
    \caption{Figure adapted from \citealt{2025ApJ...994..184C}, showing the predicted \ce{SO2} abundance as a function of planet irradiation temperature and atmospheric metallicity. The background color indicates the modeled \ce{SO2} volume mixing ratio (VMR), with darker regions corresponding to higher \ce{SO2} abundances. The horizontal cyan line marks the atmospheric metallicity inferred from our equilibrium-chemistry retrieval for LTT 9779 b. At the planet's irradiation temperature, the cyan circle highlights the region of the background color scale corresponding to the \ce{SO2} abundance inferred from our free-chemistry retrieval.}
    \label{fig:so2_shoreline}
\end{figure}

To place our \ce{SO2} abundance constraint from the free-chemistry retrieval in the context of the high metallicity favored by the equilibrium-chemistry retrieval, we compare our result to the \ce{SO2} shoreline from \citet{2025ApJ...994..184C}, which was recently revisited by \citet{2026arXiv260909893D}. We obtain a 95\% upper limit of $\log_{10}(\mathrm{VMR}) < -1.91$, with the posterior centered at a $\log_{10}(\mathrm{VMR})$ of roughly $-4.246$. We emphasize that this value does not represent a statistically significant detection of \ce{SO2}. At the high equilibrium temperature of LTT 9779 b, the posterior-preferred \ce{SO2} abundance falls within the region of the \ce{SO2} shoreline \citep{2025ApJ...994..184C} associated with relatively metal-rich atmospheres. Lower-metallicity atmospheres are generally expected to produce lower \ce{SO2} abundances. The free-chemistry retrieval therefore provides a check on the equilibrium-chemistry result. It does not independently require a high metallicity, but its preferred \ce{SO2} abundance is compatible with the high-metallicity solutions identified in the equilibrium-chemistry retrieval. Figure \ref{fig:so2_shoreline} shows the location of LTT 9779 b in this parameter space using the retrieved metallicity and \ce{SO2} abundance.

Taken together, the retrieval and cloud-modeling results point toward a common interpretation of the panchromatic transmission spectrum. The relative flatness of the spectrum does not imply the absence of an atmosphere or necessarily the absence of molecular absorbers. Instead, the observations are consistent with an atmosphere in which clouds, temperature, metallicity, and cloud fraction are sufficiently degenerate that several physically distinct atmospheric structures can reproduce the data. High-metallicity atmospheres with widespread cloud coverage remain viable and are consistent with previous work, but lower-metallicity atmospheres with patchier clouds are also allowed. Measurements in the mid-infrared would help determine whether silicate clouds are responsible for the planet's high reflectivity and its muted transmission spectrum, and constrain the composition of the silicate clouds. These measurements would provide tighter constraints on the atmospheric metallicity and composition of this unusual planet in the hot Neptune desert.

\section{Summary}\label{sec:summary}

We present a panchromatic JWST NIRISS/SOSS and NIRSpec/G395H transmission spectrum for LTT 9779 b, a resident of the hot Neptune desert. By completing a re-reduction of the NIRISS data, a new reduction of the NIRSpec data, and combining atmospheric retrievals and cloud forward modeling, we find:

\begin{itemize}

    \item The transmission spectrum exhibits muted molecular features, and our free-chemistry retrievals do not yield statistically significant detections of individual molecular absorbers. The retrievals nevertheless place constraints on several molecular abundances and favor a low terminator temperature. However, the very low retrieved temperatures may partly reflect degeneracies with atmospheric scale height and composition rather than a physically cold terminator.

    \item Clear and cloudy free-chemistry retrievals have comparable Bayesian evidence, so the transmission spectrum alone does not statistically require clouds. However, cloudy solutions remain physically well motivated and can naturally reproduce the weak spectral modulation. The inferred atmospheric metallicity is strongly dependent on cloud coverage: patchy-cloud solutions favor lower metallicities, while the 100\% cloud-coverage retrieval favors a high-metallicity branch of the posterior.

    \item The retrieved \ce{SO2} posterior provides an additional consistency check on the high-metallicity solutions permitted by the equilibrium-chemistry retrieval, although \ce{SO2} itself is not significantly detected.

    \item Our cloud forward modeling prefers small, sub-micron cloud particles, low particle column densities, high eddy diffusion coefficients, and moderate-to-high sedimentation efficiencies, corresponding to less vertically extended clouds than suggested by some previous albedo measurements. This difference may reflect the distinct atmospheric regions probed by the observations, since reflected-light measurements are primarily sensitive to dayside clouds, while transmission spectroscopy probes the terminator.

\end{itemize}

\begin{acknowledgments}
This work is based on observations made with the NASA/ESA/CSA James Webb Space Telescope. The data were obtained from the Mikulski Archive for Space Telescopes at the Space Telescope Science Institute, which is operated by the Association of Universities for Research in Astronomy, Inc., under NASA contract NAS 5-03127 for JWST. These observations are associated with programs \#1201 and \#3231. 
Support for US investigators in program \#3231 was provided by NASA through a grant from the Space Telescope Science Institute, which is operated by the Association of Universities for Research in Astronomy, Inc., under NASA contract NAS 5–26555.

We would like to thank the UNM Center for Advanced Research Computing, supported in part by the National Science Foundation, for providing the high-performance computing resources used in this work. 

This research is based upon work supported by a New Mexico Space Grant Consortium Graduate Student Fellowship through the National Aeronautics and Space Administration under NASA Cooperative Agreement No. NM-80NSSC25M7069. 

This work was performed in part at Aspen Center for Physics, which is supported by National Science Foundation grant PHY-2210452.

T.D. was partially supported by the McDonnell Center for the Space Sciences at Washington University in St. Louis.

This research made use of \texttt{nep-des} (available in \url{https://github.com/castro-gzlz/nep-des}). 

This research has made use of the NASA Exoplanet Archive \citep{2025PSJ.....6..186C}, which is operated by the California Institute of Technology, under contract with the National Aeronautics and Space Administration under the Exoplanet Exploration Program. 

\end{acknowledgments}

\facilities{JWST: NIRISS \citep{albert2023near, doyon2023near}, NIRSpec \citep{birkmann2022, jakobsen2022}; Exoplanet Archive \citep{2025PSJ.....6..186C}
}

\software{\texttt{astraeus} \citep{astraeus}, \texttt{Astropy} \citep{2013A&A...558A..33A, astropy2018, 2022ApJ...935..167A}, \texttt{batman} \citep{Kreidberg2015}, \texttt{crds} \citep{crds}, \texttt{emcee} \citep{Foreman-Mackey2013}, \texttt{Eureka!} \citep{eureka}, \texttt{ExoTiC-LD} \citep{2024JOSS....9.6816G}, \texttt{h5py} \citep{h5py}, \texttt{jwst} \citep{2023zndo...8247246B}, \texttt{Matplotlib} \citep{matplotlib}, \texttt{Multinest} \citep{2008MNRAS.384..449F, 2009MNRAS.398.1601F, 2019OJAp....2E..10F}, \texttt{numpy} \citep{numpy}, \texttt{pandas} \citep{mckinney-proc-scipy-2010}, \texttt{petitRADTRANS} \citep{2019A&A...627A..67M, 2024JOSS....9.7028B, 2024JOSS....9.5875N}, \texttt{picaso} \citep{2019ApJ...878...70B}, \texttt{scipy} \citep{scipy}, \texttt{xarray} \citep{hoyer2017xarray, hoyer_stephan_2022_7195919}, \texttt{virga} \citep{2025ApJ...994..116M, 2026AJ....171...98B}
}


\clearpage{}
\appendix
\section{Additional Data and Plots}

\subsection{NIRSpec Limb-Darkening Coefficient Comparison}\label{subsec:LD_comparison}

\begin{figure*}[htb!]
    \centering{}
    \includegraphics[width=0.9\linewidth, trim={0cm 0cm 0cm 0cm}, clip]{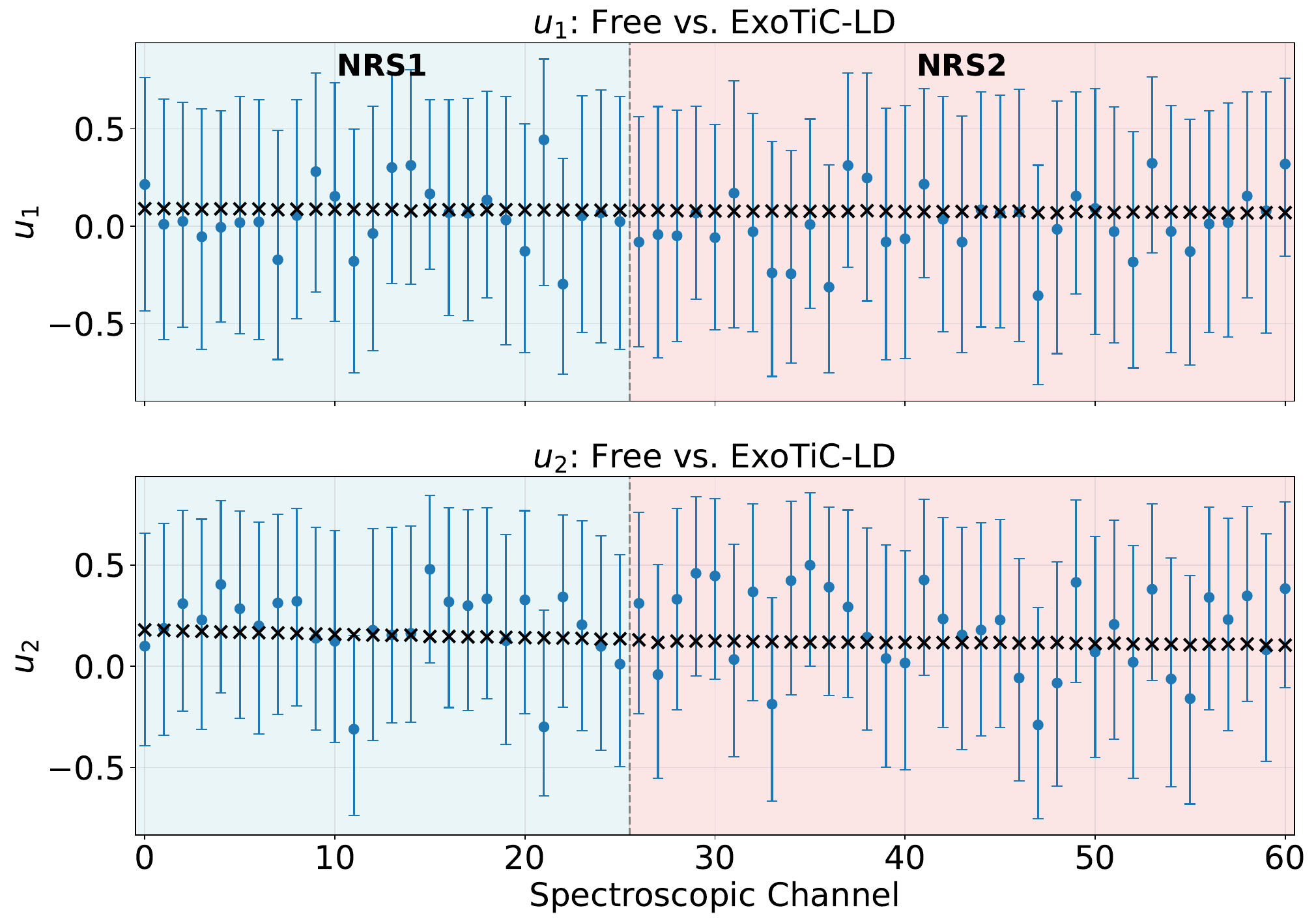}
    \caption{Comparison of quadratic limb-darkening coefficients obtained using two methods. The x's show the wavelength-dependent coefficients calculated with \texttt{ExoTiC-LD}, while the points and corresponding error bars show coefficients allowed to vary freely in the light-curve fits. The upper and lower panels show $u_1$ and $u_2$, respectively, and the shaded regions indicate the NRS1 and NRS2 detector channels.}
    \label{fig:LD_Coeff_Comparison}
\end{figure*}

\begin{figure*}[t!]
    \centering
    \includegraphics[width=0.9\linewidth, trim={0cm 0cm 0cm 0cm}, clip]
    {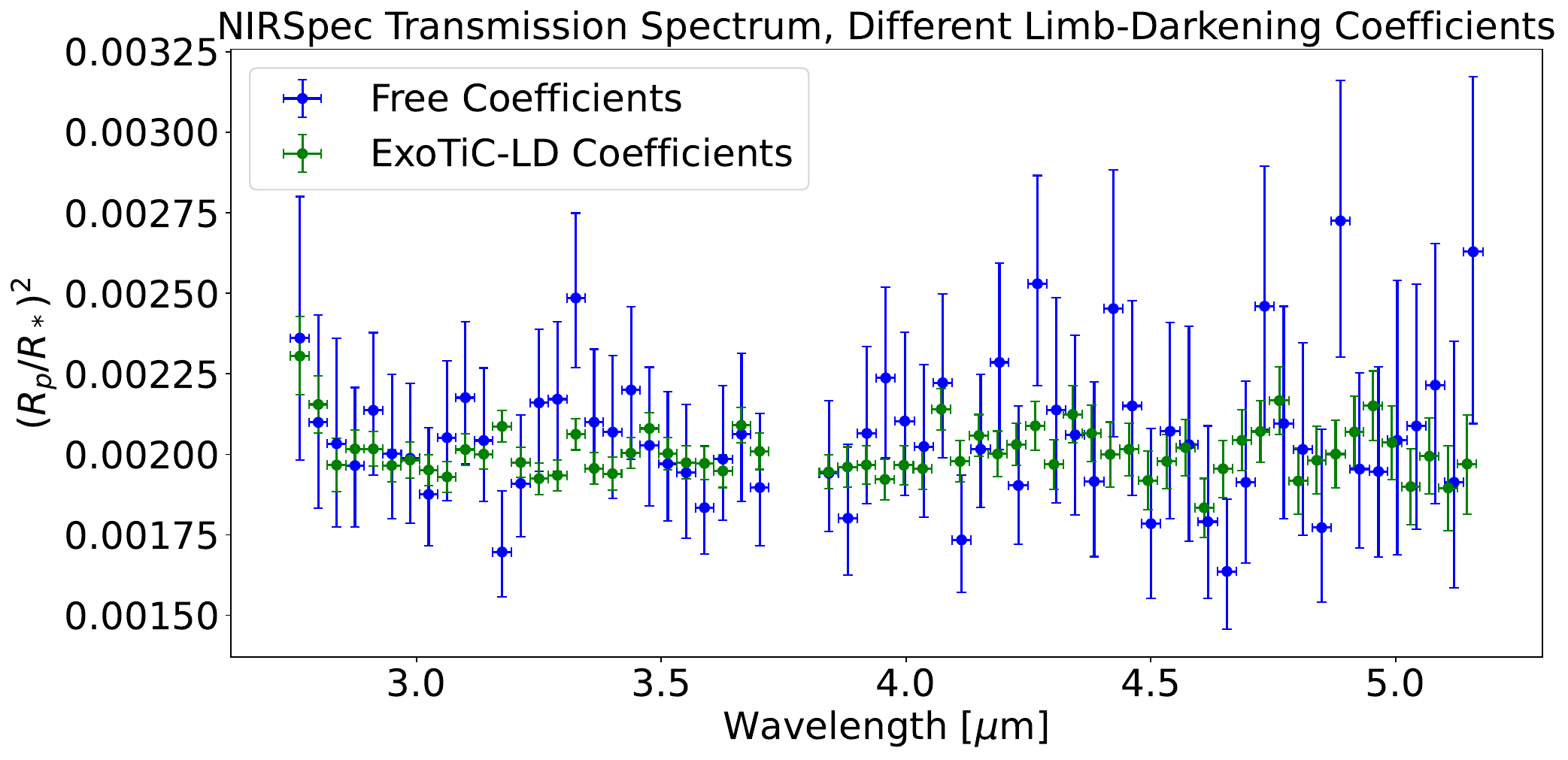}
    \caption{Comparison of two NIRSpec/G395H transmission spectra binned at approximately half of the final spectral resolution. The blue spectrum was obtained using freely fitted quadratic limb-darkening coefficients, while the green spectrum uses coefficients calculated with \texttt{ExoTiC-LD}.}
    \label{fig:LD_Coeff_Spectrum}
\end{figure*}

To calculate the free limb-darkening coefficients, we use the standard quadratic limb-darkening parameterization. Given the TESS values of $q_1$ and $q_2$ from \cite{2020NatAs...4.1148J}, we use the parameterization from \cite{2013MNRAS.435.2152K} to solve for the initial values of $u_1$ and $u_2$ for use in the light-curve fitting ($u_1$ = 0.5769, $u_2$ = 0.09391). We use a wide, uninformative prior of [-1, 1] \citep{2024AJ....168..227C} for both coefficients. The $u_1$ and $u_2$ values per channel for the free and \texttt{ExoTiC-LD} methods are shown in Figure \ref{fig:LD_Coeff_Comparison}, and NIRSpec transmission spectra using the two limb-darkening coefficient methods are shown in Figure \ref{fig:LD_Coeff_Spectrum}.

\FloatBarrier

\subsection{NIRISS Spectral Comparison to Radica et al. (2024)}
\label{subsec:NIRISS_vs_Radica}

For this work, we re-reduced the NIRISS full-orbit phase-curve data from JWST GTO 1201 with \texttt{Eureka!} as the data reduction pipeline, extracting the transmission spectrum. Figure \ref{fig:NIRISS_Spectrum} shows a comparison between this re-reduced data and the data previously shown in \cite{radica2024muted}, which was reduced with \texttt{exoTEDRF} \citep{2024JOSS....9.6898R}.

\begin{center}
    \includegraphics[width=0.9\textwidth]{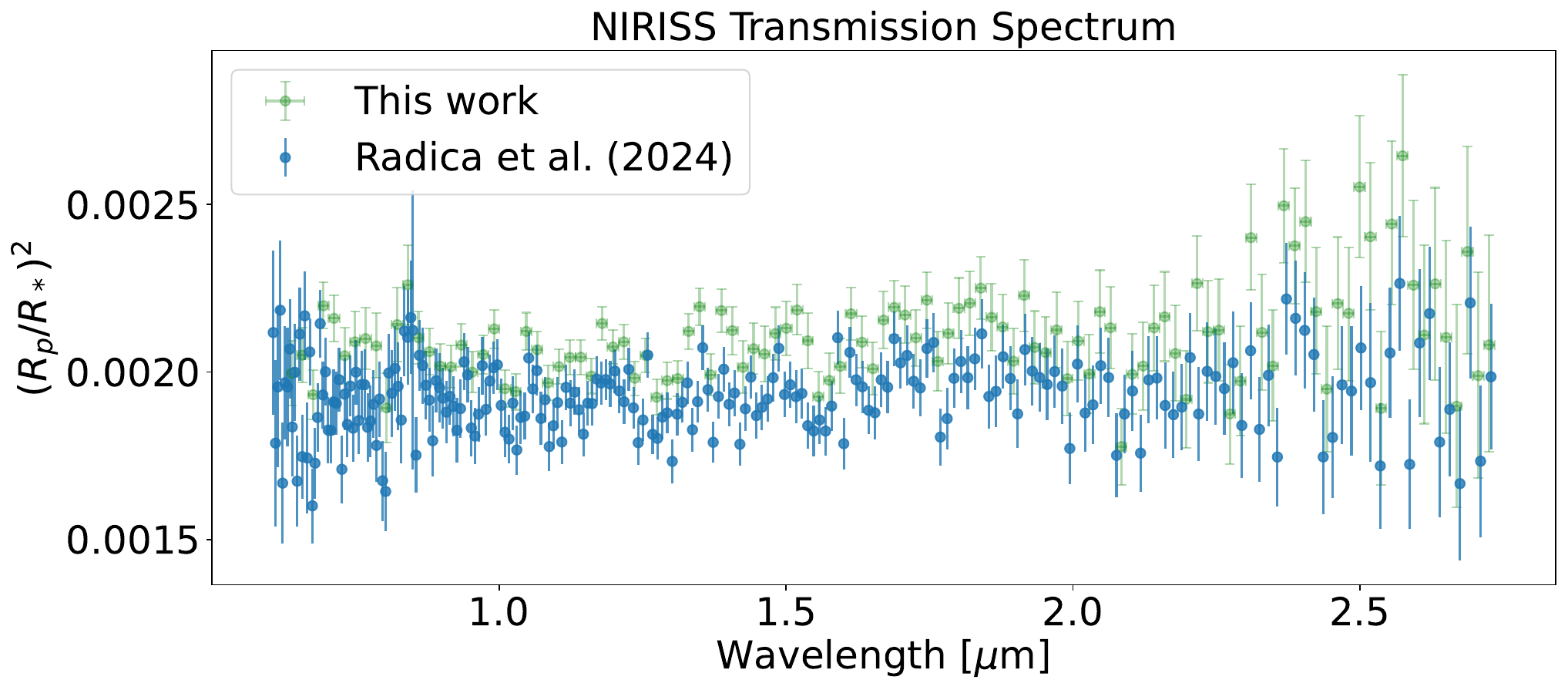}
    \captionof{figure}{Comparison of two JWST/NIRISS SOSS transmission spectra of LTT 9779 b. The green spectrum shows the \texttt{Eureka!} re-reduction presented in this work, while the blue spectrum shows the independent \texttt{exoTEDRF} reduction from \citealt{radica2024muted}.}
    \label{fig:NIRISS_Spectrum}
\end{center}

\subsection{NIRISS Spectral Comparison to HST (Edwards et al. 2023)}
\begin{figure*}[!htb]
    \centering{}
    \includegraphics[width=0.9\linewidth, trim={0cm 0cm 0cm 0cm}, clip]{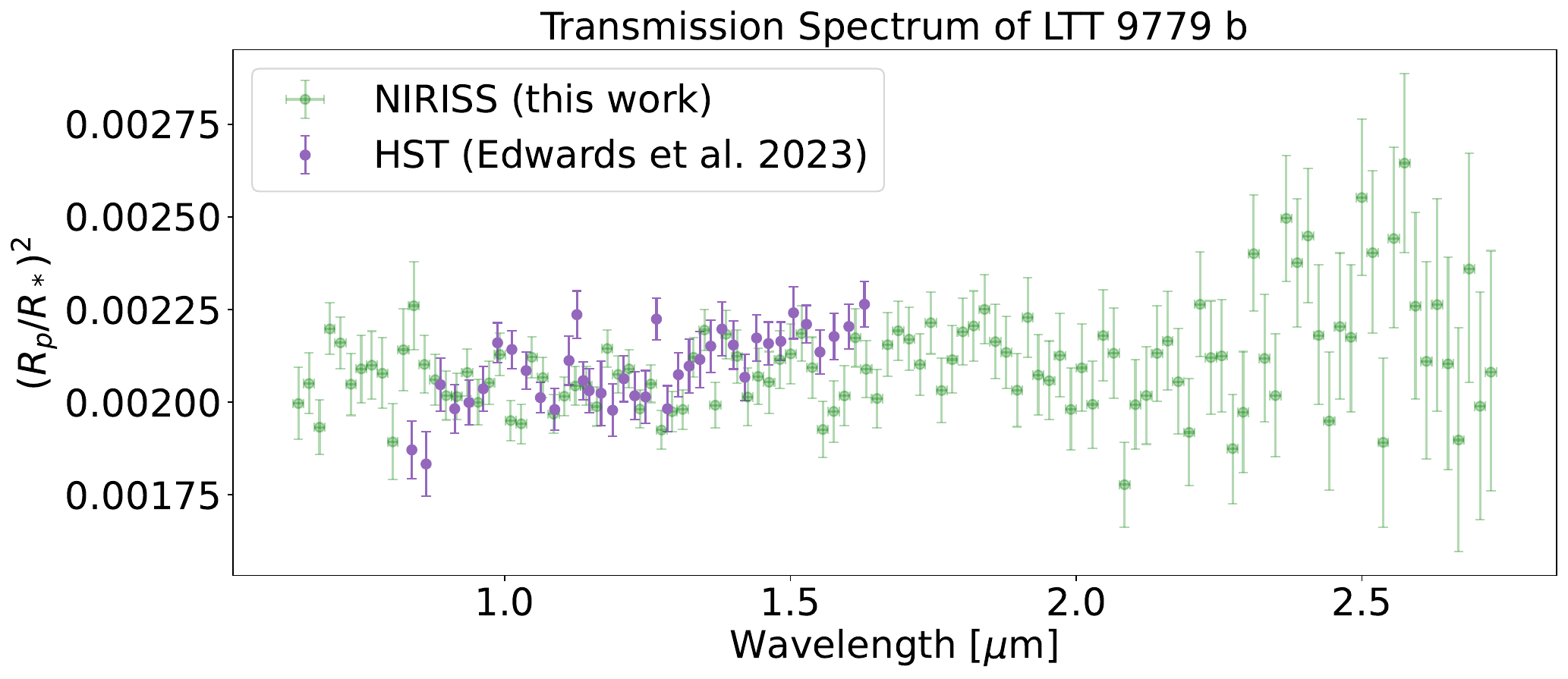}
    \caption{Comparison of the JWST/NIRISS transmission spectrum from this work and the HST transmission spectrum from \citealt{2023AJ....166..158E}. A vertical offset of 0.000061, calculated as the NIRISS median minus the HST median, was applied to the HST data for visual comparison.}
    \label{fig:NIRISS_Spectrum_with_HST}
\end{figure*}

As a further comparison point and verification of the NIRISS reduction using \texttt{Eureka!}, we looked at previous HST transmission observations, which led to an atmospheric model with evidence for \ce{H2O} and \ce{CO2}, as well as a non-detection of atmospheric escape \citep{2023AJ....166..158E}. Figure \ref{fig:NIRISS_Spectrum_with_HST} shows the NIRISS spectrum from this work compared to the HST transmission spectrum.

\subsection{Clear Atmospheric Retrieval}\label{subsec:clearretrieval}
\begin{figure*}[!htb]
    \centering{}
    \includegraphics[width=1.0\linewidth, trim={0cm 0cm 0cm 0cm}, clip]{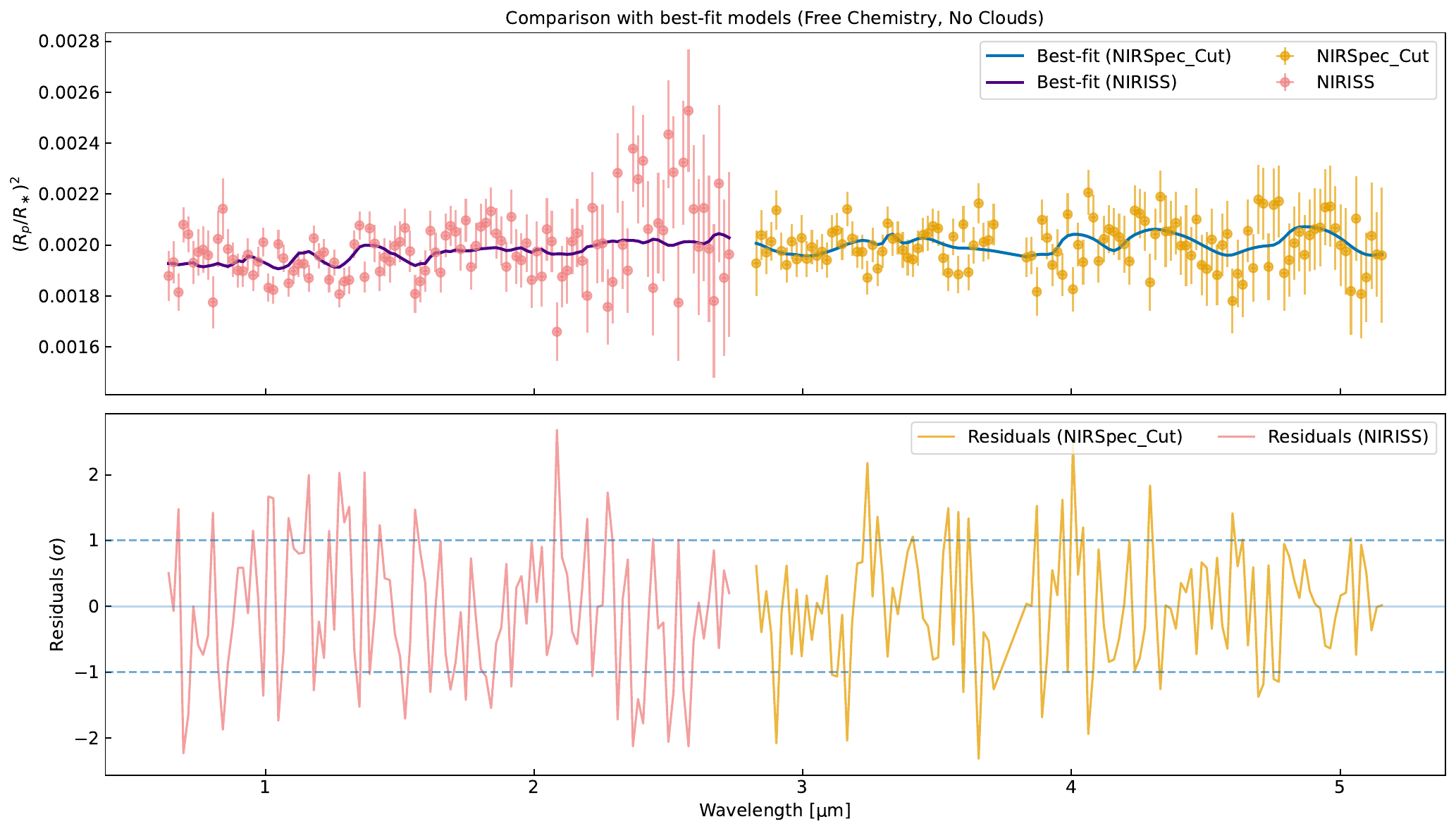}
    \caption{Panchromatic NIRISS/NIRSpec transmission spectrum with the best-fit clear free-chemistry retrieval overplotted. The upper panel shows the observed transit depths and corresponding model spectra for NIRISS and NIRSpec, while the lower panel shows the residuals in units of the observational uncertainty.}
    \label{fig:free_clear_spectrum}
\end{figure*}

\begin{figure*}[!htb]
    \centering{}
    \includegraphics[width=1.0\linewidth, trim={0cm 0cm 0cm 0cm}, clip]{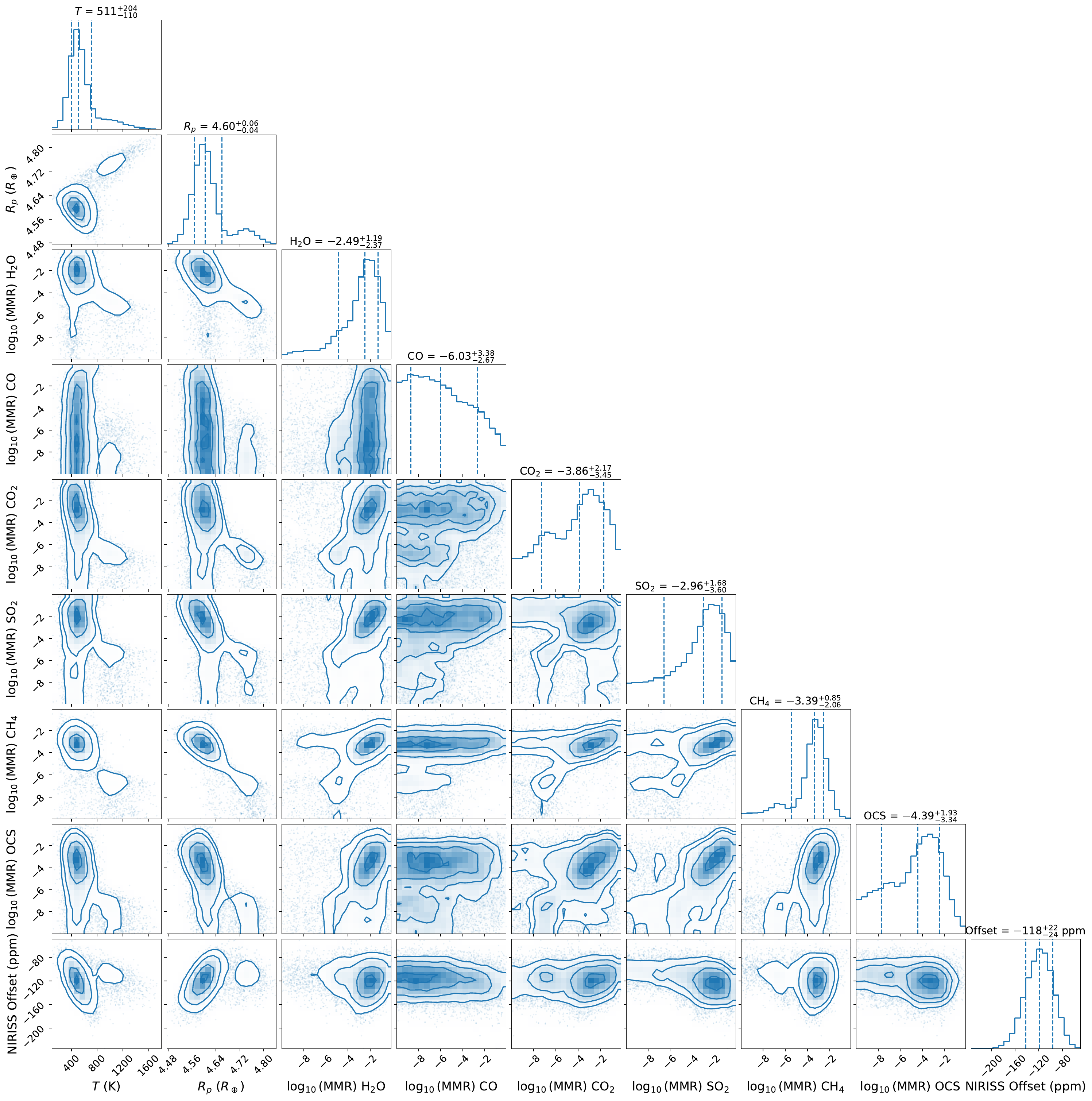}
    \caption{Posterior distributions for the clear free-chemistry retrieval. Retrieved parameters include the isothermal atmospheric temperature, planetary radius, molecular abundances of \ce{H2O}, \ce{CO}, \ce{CO2}, \ce{SO2}, \ce{CH4}, and \ce{OCS}, and the NIRISS/NIRSpec offset. Molecular abundances are reported as $\log_{10}$ mass mixing ratios (MMRs), and the NIRISS/NIRSpec offset is shown in ppm.}
    \label{fig:free_clear_cornerplot}
\end{figure*}

Results from the free chemistry atmospheric retrieval without clouds are shown in Figures \ref{fig:free_clear_spectrum} and \ref{fig:free_clear_cornerplot}.

\FloatBarrier

\subsection{Additional Cloudy Atmospheric Retrievals}
\label{subsec:cloudy_retrieval_posteriors}

\begin{figure*}[!htb]
    \centering{}
    \includegraphics[
        width=1.0\linewidth,
        trim={0cm 0cm 0cm 0cm},
        clip
    ]{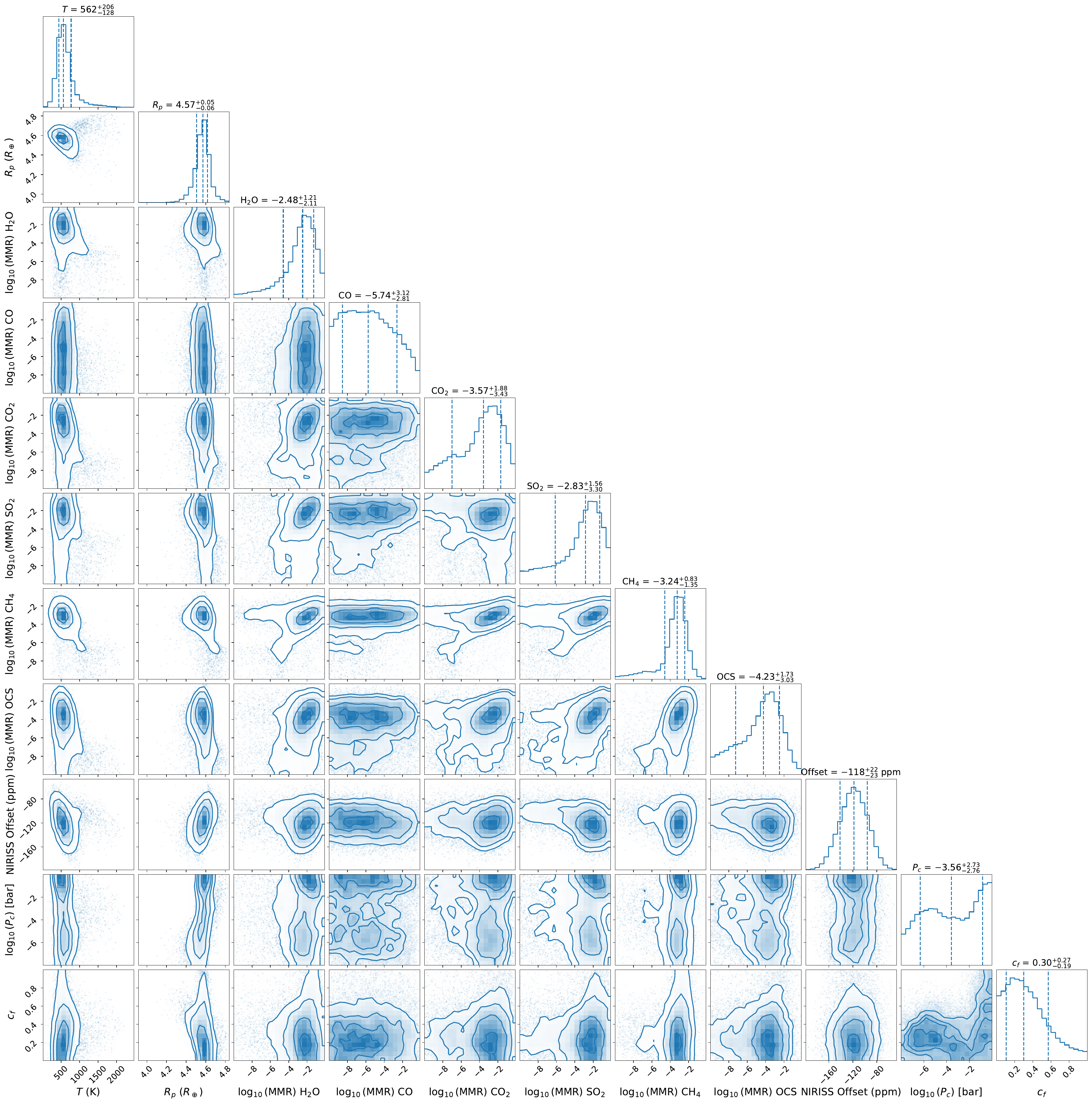}
    \caption{Posterior distributions for the patchy-cloud free-chemistry retrieval. Retrieved parameters include the isothermal atmospheric temperature, planetary radius, molecular abundances of \ce{H2O}, \ce{CO}, \ce{CO2}, \ce{SO2}, \ce{CH4}, and \ce{OCS}, the NIRISS/NIRSpec offset, opaque cloud-top pressure, and cloud fraction. Molecular abundances are reported as $\log_{10}$ mass mixing ratios (MMRs), and the NIRISS/NIRSpec offset is shown in ppm.}
    \label{fig:free_clouds_cornerplot}
\end{figure*}

\begin{figure*}[!htb]
    \centering{}
    \includegraphics[
        width=1.0\linewidth,
        trim={0cm 0cm 0cm 0cm},
        clip
    ]{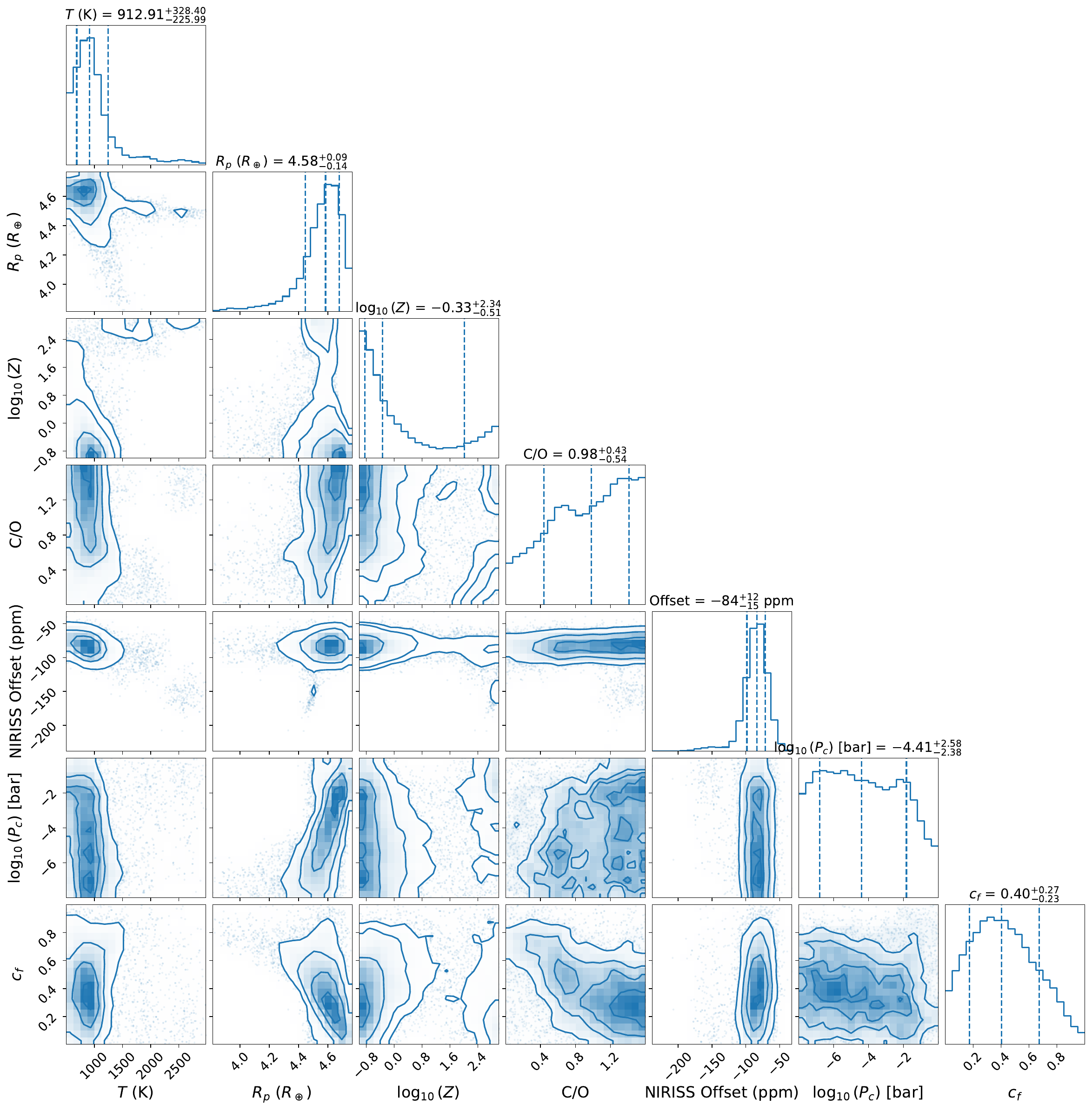}
    \caption{Posterior distributions for the patchy-cloud equilibrium-chemistry retrieval. Retrieved parameters include the isothermal atmospheric temperature, planetary radius, atmospheric metallicity, C/O ratio, NIRISS/NIRSpec offset, opaque cloud-top pressure, and cloud fraction.}
    \label{fig:eqm_clouds_cornerplot}
\end{figure*}

The posterior distributions for the patchy-cloud free-chemistry and patchy-cloud equilibrium-chemistry retrievals are shown in Figures \ref{fig:free_clouds_cornerplot} and \ref{fig:eqm_clouds_cornerplot}, respectively. The corresponding best-fit spectra are shown together with the 100\% cloud-fraction equilibrium-chemistry retrieval in Figure \ref{fig:retrieval_comparison} in the main text.

\FloatBarrier

\subsection{Bayesian Model Comparison Results}\label{subsec:leaveoneout}
Table \ref{tab:significances} shows the free-chemistry retrievals run, the $\mathrm{ln}(Z)$ evidence values, and the $\Delta \mathrm{ln}(Z)$ for each case. The $\Delta \mathrm{ln}(Z)$ values are small, with some negative values present as well. This indicates that the current transmission spectrum does not require the inclusion of \ce{CO}, \ce{CO2}, \ce{H2O}, \ce{CH4}, \ce{SO2}, or \ce{OCS} to explain the data. This suggests that any molecular features are likely muted, consistent with a cloudy or low-amplitude transmission spectrum. There is also a slight preference for the cloud-free model versus the clear model.

\begin{table}[htbp]
\centering
\caption{Free chemistry retrievals run and evidence values for different cloudy/clear and molecular combinations.}
\label{tab:significances}
\begin{tabular}{lrrr}
\toprule
Retrieval Details & $\mathrm{ln}\,Z$ & $\Delta \mathrm{ln}\,Z$ & $\sigma$ \\ 
\midrule
Cloudy & -130.5 & 0.00 & \dots \\
Clear & -129.7 & -0.8 & \dots \\
Cloudy, Leave Out \ce{CO} & -129.9 & -0.6 & \dots \\
Cloudy, Leave Out \ce{CO2} & -129.3 & -1.2 & \dots \\
Cloudy, Leave Out \ce{CH4} & -131.3 & 0.8 & \dots \\
Cloudy, Leave Out \ce{SO2} & -130.7 & 0.2 & \dots \\
Cloudy, Leave Out \ce{H2O} & -131.3 & 0.8 & \dots \\
Cloudy, Leave Out \ce{OCS} & -130.4 & -0.1 & \dots \\
\bottomrule
\end{tabular}
\end{table}

\FloatBarrier

\subsection{Cloud Forward Model Condensate Comparison}\label{subsec:cloud_condensates}
\begin{figure*}[!htb]
    \centering{}
    \includegraphics[width=0.9\linewidth, trim={0cm 0cm 0cm 0cm}, clip]{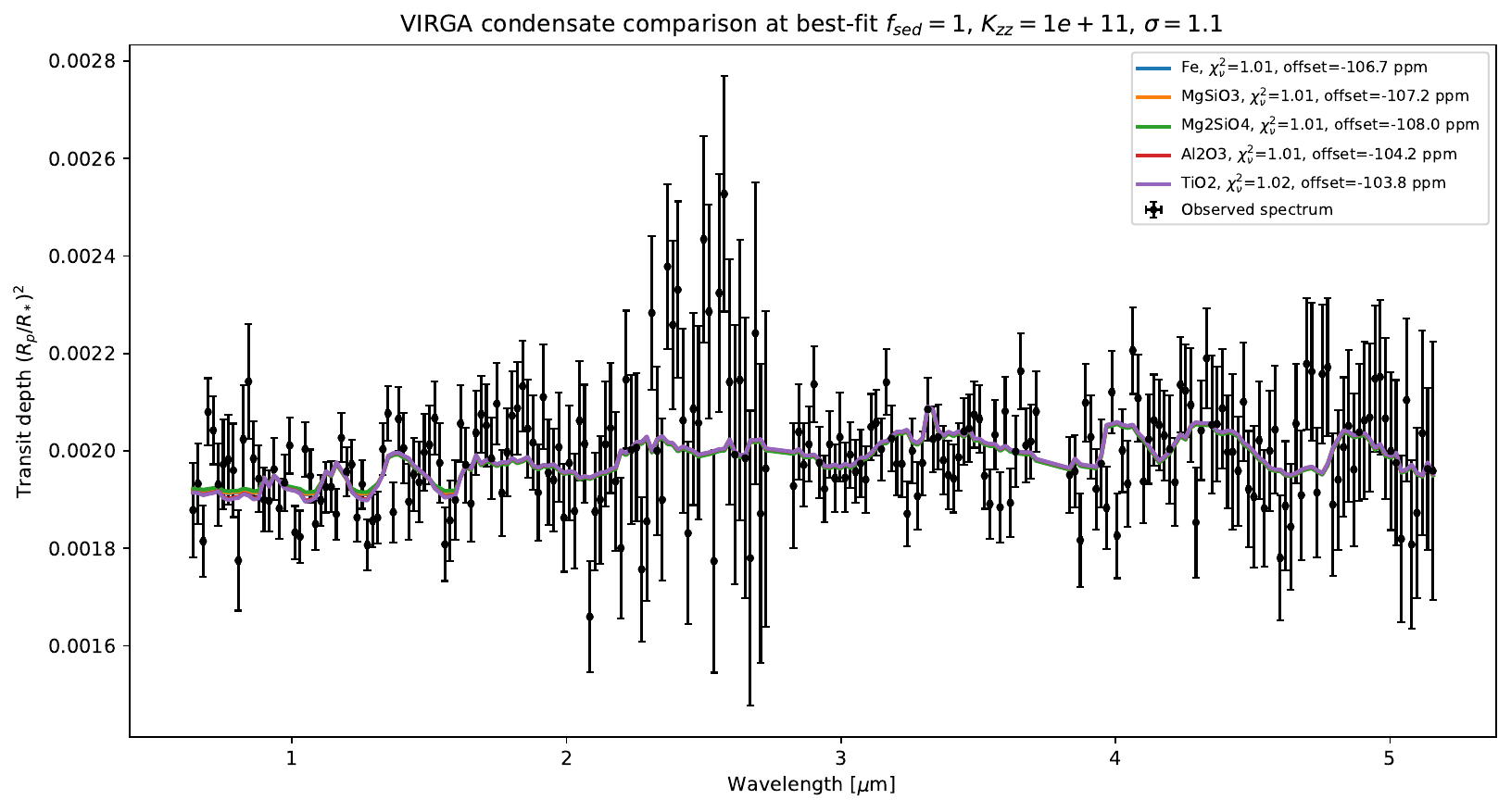}
    \caption{Comparison of candidate cloud condensates in the \texttt{virga}+\texttt{picaso} forward-model framework at fixed $f_{\rm sed}=1$, $K_{zz}=10^{11}$~cm$^2$~s$^{-1}$, and particle-size distribution width $\sigma=1.1$. Models for \ce{Fe}, \ce{MgSiO3}, \ce{Mg2SiO4}, \ce{Al2O3}, and \ce{TiO2} are compared with the observed panchromatic transmission spectrum. The legend reports the reduced $\chi^2$ and fitted vertical offset for each condensate model.}
    \label{fig:cloud_species_comparison}
\end{figure*}

Using a fixed $f_{\rm sed}$ value, we compare five different condensates to determine if we can distinguish between the condensates with the panchromatic transmission spectrum. At fixed $f_{\rm sed}$ = 1, the condensate models produce nearly indistinguishable spectra and comparable reduced $\chi^2$ values, indicating that the current data do not strongly distinguish between the tested cloud species. These results are shown in Figure \ref{fig:cloud_species_comparison}.

\FloatBarrier

\FloatBarrier

\subsection{Zoom-In on Independent Atmospheric Retrievals}\label{subsec:independent_retrieval_zoomin}
\FloatBarrier

\begin{figure}[h!]
\centering
    \subfloat{\includegraphics[width=0.8\linewidth]{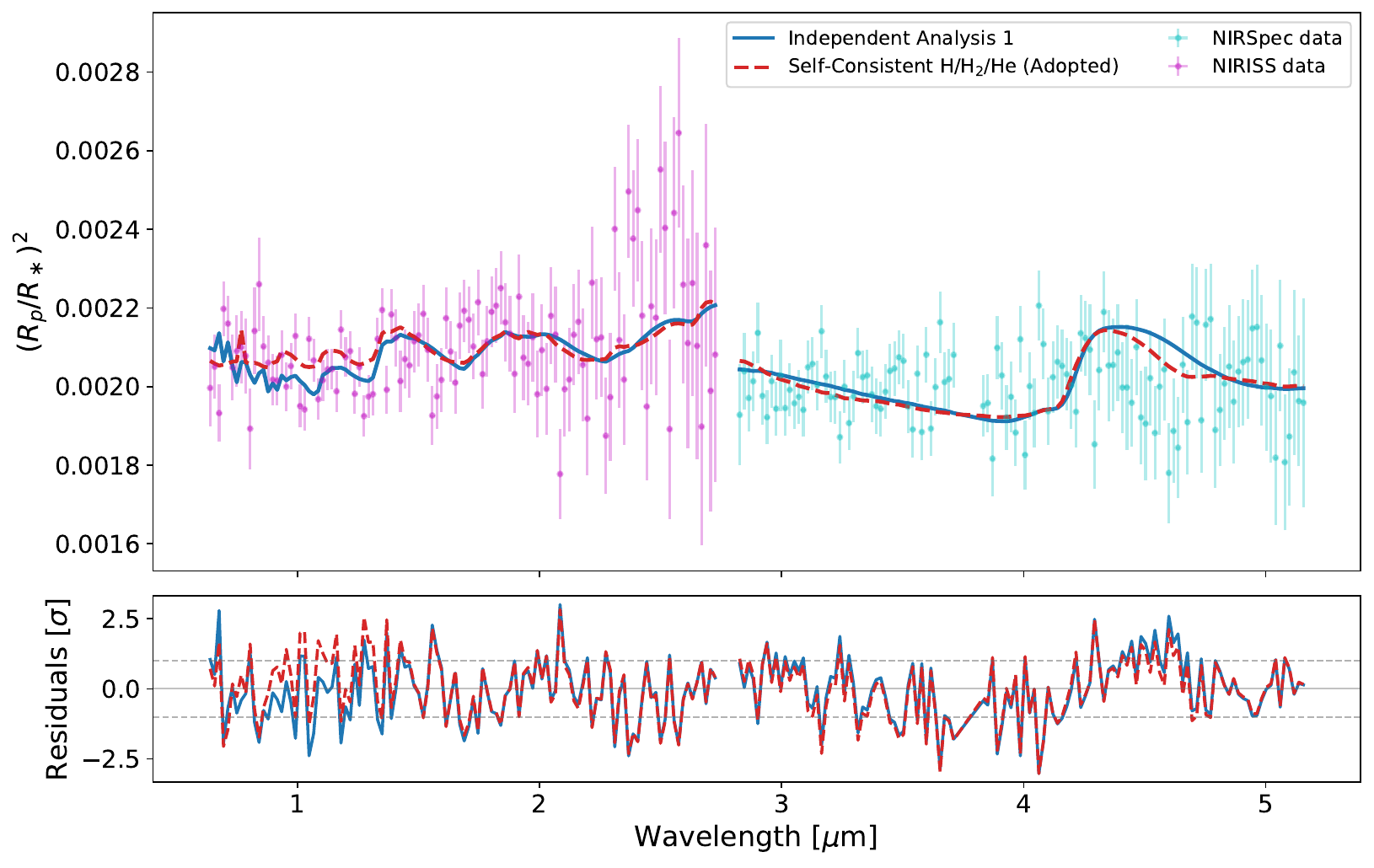}}\quad
    \subfloat{\includegraphics[width=0.8\linewidth]{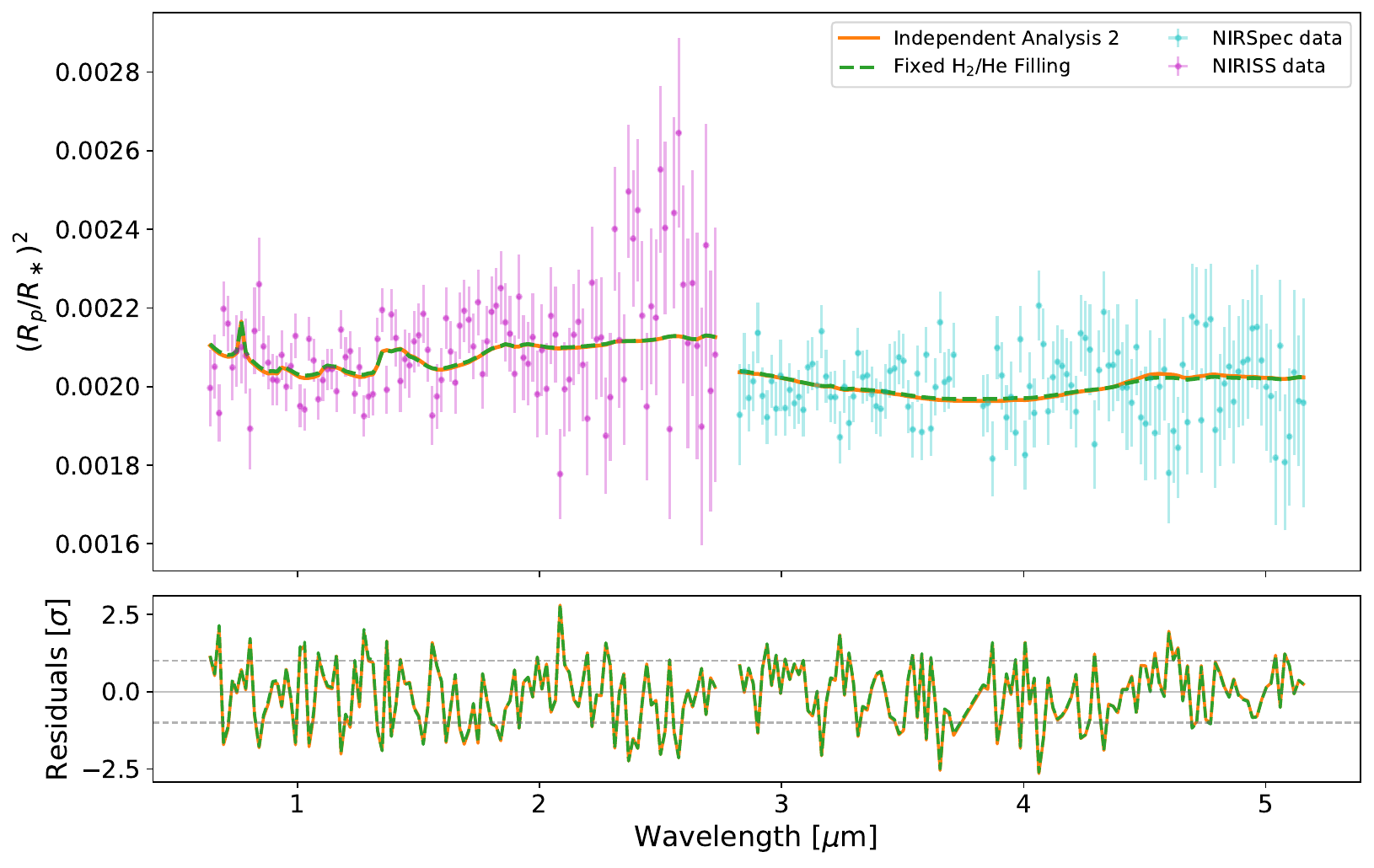}}
    \caption{Detailed comparison of equilibrium-chemistry retrievals illustrating the effect of the \ce{H2}/\ce{He} filling prescription. The upper comparison shows Independent Analysis 1 alongside the adopted self-consistent treatment, in which the \ce{H}, \ce{H2}, and \ce{He} abundances are determined by the equilibrium-chemistry calculation. The lower comparison shows Independent Analysis 2 alongside the earlier fixed-filling implementation, in which the remaining atmospheric mass fraction is assigned to \ce{H2} and \ce{He} in a fixed ratio. NIRISS and NIRSpec data are shown in both panels, with residuals in units of the observational uncertainty below each spectral comparison.}\label{fig:detailed_eqm_independent_analysis}
\end{figure}

\FloatBarrier
\clearpage{}

\bibliography{ltt9779b}{}
\bibliographystyle{aasjournalv7}


\end{document}